\documentclass[useAMS,usenatbib]{mn2e}

\usepackage{verbatim,graphicx,epsfig,dcolumn}
\newcolumntype{.}{D{.}{.}{4}}
\newcolumntype{,}{D{.}{.}{2}}
\newcolumntype{;}{D{.}{.}{1}}
\newcommand{\nodata}{$\cdot\cdot\cdot$}
\newcommand{\lesssim}{{\lower-1.2pt\vbox{\hbox{\rlap{$<$}\lower5pt\vbox{\hbox{$\sim$}}}}}}
\newcommand{\gtrsim}{{\lower-1.2pt\vbox{\hbox{\rlap{$>$}\lower5pt\vbox{\hbox{$\sim$}}}}}}
\usepackage{color}
\definecolor{red}{rgb}{0.75,0.0,0.0}

\title[VISTA's view of the Sgr dSph and Bulge]{VISTA's view of the Sagittarius dwarf spheroidal galaxy and southern Galactic Bulge}
\author[I. McDonald, et al.]{I.~McDonald$^{1}$\thanks{E-mail: mcdonald@jb.man.ac.uk}, A.~A.~Zijlstra$^{1}$, G.~C.~Sloan$^{2}$, E.~J.~Kerins$^{1}$, E.~Lagadec$^{2}$,
\newauthor D.~Minniti$^{3,4}$, M.~V.~Santucho$^{5,6}$, S.~Gurovich$^{5,6}$,
\newauthor M.~J.~de~L. Dom\'inguez Romero$^{5,6}$\\
$^{1}$Jodrell Bank Centre for Astrophysics, Alan Turing Building, Manchester, M13 9PL, UK\\
$^{2}$Cornell University, Astronomy Department, Ithaca, NY 14853-6801, USA\\
$^{3}$Departamento de Astronomia y Astrofisica, Pontificia Universidad Cat\'olica de Chile, Vicu\~na Mackenna 4860, Casilla 306,\\ \, Santiago 22, Chile\\
$^{4}$Vatican Observatory, V00120 Vatican City State, Italy\\
$^{5}$Instituto de Astronom\'ia Te\'iorica y Experimental (IATE-CONICET), Laprida 922 X5000BGR C\'ordoba, Argentina\\
$^{6}$Observatorio Astron\'omico de la Universidad Nacional de C\'ordoba, Argentina}

\begin{document}

\date{Accepted 9999 December 32. Received 9999 December 32; in original form 9999 December 32}

\pagerange{\pageref{firstpage}--\pageref{lastpage}} \pubyear{9999}

\maketitle

\label{firstpage}

\begin{abstract}
We present the deepest near-infrared ($ZJK_s$) photometry yet obtained of the Sagittarius dwarf spheroidal (Sgr dSph), using VISTA to survey 11 square degrees centred on its core. We list locations and $ZJK_s$-band magnitudes for over 2.9 million sources in the field. We discuss the isolation of the Sgr dSph from the foreground and Galactic Bulge populations, identify the Sgr dSph's horizontal branch in the near-infrared for the first time, and map the density of the galaxy's stars. We present isochrones for the Sgr dSph and Bulge populations. These are consistent with the previously-reported properties of the Sgr dSph core: namely that it is dominated by a population between [Fe/H] $\approx$ --1 dex and solar, with a significant [$\alpha$/Fe] versus [Fe/H] gradient. While strong contamination from the Galactic Bulge prevents accurate measurement of the (Galactic) north side of the Sgr dSph, the dwarf galaxy can be well-approximated by a roughly ovaloid projection of characteristic size $4^\circ \times 2^\circ$, beyond which the projected stellar density is less than half that of the region surrounding the core. The galaxy's major axis is perpendicular to the Galactic Plane, as in previous studies. We find slight evidence to confirm a metallicity gradient in the Sgr dSph and use isochrones to fit a distance of 24.3 $\pm$ 2.3 kpc. We were unable to fully constrain the metallicity distribution of the Sgr dSph due to the Bulge contamination and strong correlation of [$\alpha$/Fe] with metallicity, however we find that metal-poor stars ([Fe/H] $\lesssim$ --1) make up $\lesssim$29 per cent of the Sgr dSph's upper-RGB population. The Bulge population is best fit by a younger population with [Fe/H] $\approx$ 0 and [$\alpha$/Fe] $\approx$ 0 or slightly higher. We find no evidence for a split, peanut- or X-shaped Bulge population in this line of sight ($l = 5.6^\circ \pm \sim 1^\circ$, $b = -14.1^\circ \pm \sim 3^\circ$).
\end{abstract}



\section{Introduction}
\label{IntroSect}

The histories of Local Group galaxies offer a reflection of our own Galaxy's history and of its future in the context of Universal cosmology \citep{FBH02,McConnachie12}. Tidal dissipation of satellite galaxies into their hosts, as part of ongoing hierarchical galaxy formation, directly affects the star-forming, dynamical and chemical evolution of the larger system (e.g.\ \citealt{Peebles69,GG72,EJ79,MKL+96,Ibata02,BBN+04}). Studying our nearest galactic neighbours reveals the processes of competitive galactic accretion into the Milky Way, of tidal stripping of dwarf galaxies, and of the incorporation of their stars into the heterogeneous population we see in the Solar Neighbourhood, Galactic Disc and Halo (e.g.\ \citealt{HWdZZ99,KMH+07,Klement10}).

The Sagittarius Dwarf Spheroidal (Sgr dSph; \citealt{IGI94}), is our nearest confirmed galactic neighbour at 25 kpc distance\footnote{The putative Canis Major dwarf spheroidal (\citealt{MIB+04}) would be closer still, but despite extensive and lively discussion in recent literature there remains considerable debate regarding its extragalactic nature \citep{MZB+04,MVZ+09,SVGMD+11,MFJF12}, summarised in the unpublished work of \citet{LCMZ+12}.} \citep{MKS+95,MBFP04,KC09}. We see it caught in the Galaxy's gravitational pull as it is being tidally torn apart. Its tidal streams comprise 70\% of its total light \citep{NOBEP10} and span the entire sky, tracing multiple orbits around the Galaxy \citep{MSWO03,KBE+12,SBS+13}. These streams are a primary tool in modelling the mass distribution of the Milky Way (e.g.\ \citealt{DW13}). The original dwarf galaxy seems to have been somewhat smaller than the Small Magellanic Cloud ($\sim$10$^{10}$ M$_\odot$; \citealt{NOBE12}), but its impact into the Galaxy has nevertheless been suggested as the underlying cause of the Milky Way's current spiral structure \citep{PBT+11}.

Though well-approximated by a spheroidal shape showing little or no rotation \citep{PZI+11,MWZ+12}, the surviving part of the galaxy has multiple populations and a noteable metallicity gradient (\citealt{MUS+95,SL95,FMR+96,MMU+96,LS00,MBB+05,SDM+07}, hereafter SDM+07; \citealt{BIC+08}) and large-scale density enhancements which appear to deviate from the overall spheroid \citep{IGI94}. Historically, several Galactic globular clusters appear to have been associated with the Sgr dSph\footnote{Though authors differ on which clusters are associated, the following are generally accepted: M54, Terzan 7, Terzan 8, Arp 2, Palomar 12, Whiting 1 (\citealt{DCA95,LM10} and references therein).} and, while the cluster M54 has been suggested as the core of the Sgr dSph galaxy, it is unexpectedly metal-poor for such an object and may simply be projected onto a region of high density \citep{SML+11}.

The galaxy's star-formation history is complex, with several spectroscopically identified but poorly delineated populations. SDM+07 have summarised them into four (or possibly five) photometrically-defined groups, which we investigate in Section \ref{IsoSgrSect}. There are strong correlations between metallicity and both age and $\alpha$-element enhancement between M54 ([Fe/H] = --1.7 dex, [$\alpha$/Fe] $\approx$ +0.2 dex, $t \approx 13$ Gyr) and the youngest population around solar metallicity ([$\alpha$/Fe] $\approx$ --0.2 dex, $t \approx 2.3$ Gyr). A potential super-solar metallicity population of $t < 1$ Gyr may also exist, though the SDM+07 acknowledge that the evidence for it is not strong.

The proximity of the Sgr dSph to the Galactic centre ($l$ = 5.6, $b$ = --14.1) has hampered efforts to characterise its structure and history. A large part of the galaxy lies behind the Galactic Bulge, which remains the primary contaminant across its face. So strong is the contamination that the galaxy cannot be visually separated from the Bulge: we must rely on colour--magnitude diagrams, such as those used to discover it \citep{IGI94}, or time-consuming radial velocity or variability observations (e.g.\ \citealt{MWZ+12}, or the aforementioned RR Lyr detections) to identify its members and map its structure.

In this work, we present $ZJK_s$-band observations of the inner regions of the Sgr dSph with the European Southern Observatory's Visible and Infrared Survey Telescope for Astronomy (VISTA), which we use to create the deepest-yet infrared colour--magnitude diagram of the Sgr dSph galaxy, with which we probe its stellar populations and their physical extent in the galaxy.


\section{Observations}
\label{ObsSect}

\subsection{Observation and reduction}
\label{RedSect}

The VISTA Infrared Camera (VIRCAM) surveyed the core of the Sgr dSph over 16 nights between 2012 Apr 06 and 2012 Jun 07. VIRCAM consists of each 16 imaging chips which create a pawprint that must be moved around the sky to create a contiguous imaging tile of around 1.5 square degrees. The survey consists of seven VIRCAM tiles, covering an irregularly-shaped pattern chosen to match the densest parts of the Sgr dSph identified by \citet{IGI94}. Figure \ref{MapFig} shows the survey area. Each tile was observed once in the $J$- and $K_s$-bands, and 12 times in the $Z$-band. In this paper, we will use a time-average of those $Z$-band observations, which will be used to provide variability information in a subsequent paper.

During the exposures, VIRCAM was set in FPJME nesting mode\footnote{Full documentation covering the VISTA observing modes can be found in the User Manual: http://www.eso.org/sci/facilities/paranal/instruments/vircam/.}. In this mode, each tile is repeated for each filter, and for each filter the array is repointed such that the pawprint fills in the gaps in the tile pattern. The Tile6zz pattern was used, which follows a zig-zag pattern which ensures that the pawprints cover every region of the tile twice, building redundancy. For each pawprint, the telescope was randomly jittered by 15$^{\prime\prime}$ four times, and within each jitter point the array was micro-stepped by half a pixel each way in each axis (using the Ustep2x2 micro-stepping pattern, resulting in four microstep pointings per jitter), with two exposures per micro-step. Exposure times were set to 4 s in each channel to ensure maximum saturation of the galaxy's AGB stars without loss of linearity, giving an effective exposure time of 64 s to most of the observed area.

Following on-site quality checks and calibration on-site at Paranal and subsequently by the European Southern Observatory at Garching, data reduction to the level of aperture photometry was carried out automatically by the Cambridge Astronomy Survey Unit (CASU)\footnote{Full documentation covering the VISTA/VIRCAM pipeline can be found at: http://apm49.ast.cam.ac.uk/surveys-projects/vista/technical.}. Image processing by CASU comprises of the usual image resotration process (corrections for the detector reset, dark, linearity and flat-field functions), plus more-advanced image processing (sky background and nebulosity subtraction, de-striping, jitter and pawprint stacking). Source extraction is then performed on the stacked, tiled images by CASU, which is then astronomically and photometrically calibrated to 2MASS to generate their final catalogues.

Data retrieved from CASU therefore comprised a list of astrometrically-calibrated object detections for individual tiles and epochs in individual filters. Further analysis was required to correct the object fluxes for point-spread function distortions, merge the catalogues taken in different filters and epochs, remove objects duplicated between tiles, and systematically correct magnitudes for photometric zero points. We used magnitudes calculated within the recommended aperture \#3, which corresponds to the flux received by a circular aperture of 1$^{\prime\prime}$ radius, and followed the distortion correction routine outlined in the VISTA documentation\footnote{CASU VIRCAM notes: http://apm49.ast.cam.ac.uk/surveys-projects/vista/technical/catalogue-generation .}.

The simultaneous $ZJK_s$-band images allowed for a relatively simple merging procedure. Objects identified in different filters were simply grouped by finding the nearest $J$- and $K_s$-band neighbour to each $Z$-band detection. Objects within 0$\farcs$8 of each other were considered to be detections of the same object (cf.\ typical full-width--half-maximum point-spread function size of 1$\farcs$7, with a typical $\lesssim\pm$0$\farcs$1 variation over the image plane). In principle, we may miss exceptionally red objects this way, but the depth of the catalogue ensures that the reddest AGB stars (known already from 2MASS; \citealt{CSvD+03}) will still be identified, thus the only missing objects will be faint, red background galaxies or exceptionally cool foreground objects, neither of which are associated with the Sgr dSph. The same principle was followed to merge the multi-epoch $Z$-band images, resulting in a catalogue for each tile with 12 $Z$-band magnitudes and one $J$-band and one $K_s$-band magnitude. The median difference between the $Z$-band magnitudes at each epoch was found, and an offset applied to give each $Z$-band epoch the same zero-point magnitude, though we note that this has not yet been absolutely calibrated. We discuss this in the Appendix.

Tile merging to form a master catalogue followed a similar principle. Detections in different tiles were flagged as the same object when they fell within 1$\farcs$2 arcseconds of each other. This was done iteratively, as some objects were observed in up to four tiles, thus we have up to four independent detections of the same object in some cases. The source tile number was retained during this process to allow a per-tile zero-point correction later.

We can, to first order, check the effectiveness of the source matching by comparing the source densities in regions where the tiles overlap to regions covered by only one tile. If the matching is too liberal, then the overlap regions will be under-dense in sources; if it is too conservative, they will remain over-dense. The top panel of Figure \ref{MapFig} shows that the tile overlap areas have the same source density as the centres of the tiles, thus the matching process is effective to first order.

The same figure highlights considerable differences in source density between tiles. This originates in the original CASU pipeline reduction, and is a result of differing signal-to-noise limits between tiles, which were taken on different epochs with different seeing conditions and different levels of lunar and other background light. When we limit the data to only bright sources (bottom panel of Figure \ref{MapFig}), the discrepencies between tiles become smoother and the tiling pattern disappears, indicating that our completeness is similar across all tiles down to the magnitude of the Sgr dSph horizontal branch ($K_{\rm s} \sim 16$).

An identical algorithm to the tile merging processing was followed for matching literature catalogues, again with a 1$\farcs$2 arcsecond maximum offset. The merged catalogue was cross-correlated with the 2MASS, UCAC4 (Zacharias et al., in prep.) and \emph{WISE} \citep{CWC+12} catalogues to give a multi-mission catalogue spanning from the optical ($B$-band) to mid-infrared (22 $\mu$m).

The final merged catalogue was then calibrated to the 2MASS Vega-based photometric zero-point system, taking into account the difference in filter transmissions between the two and assuming a reddening of $E(B-V) = 0.15$ (valid for M54; \citealt{Harris96}) or $E(J-K) = 0.08$. Details on the source density limits, comparison to 2MASS and filter corrections are given in the Appendix.

\subsection{Final catalogue}
\label{CatSect}

\begin{figure*}
\centerline{\includegraphics[height=0.70\textwidth,angle=-90]{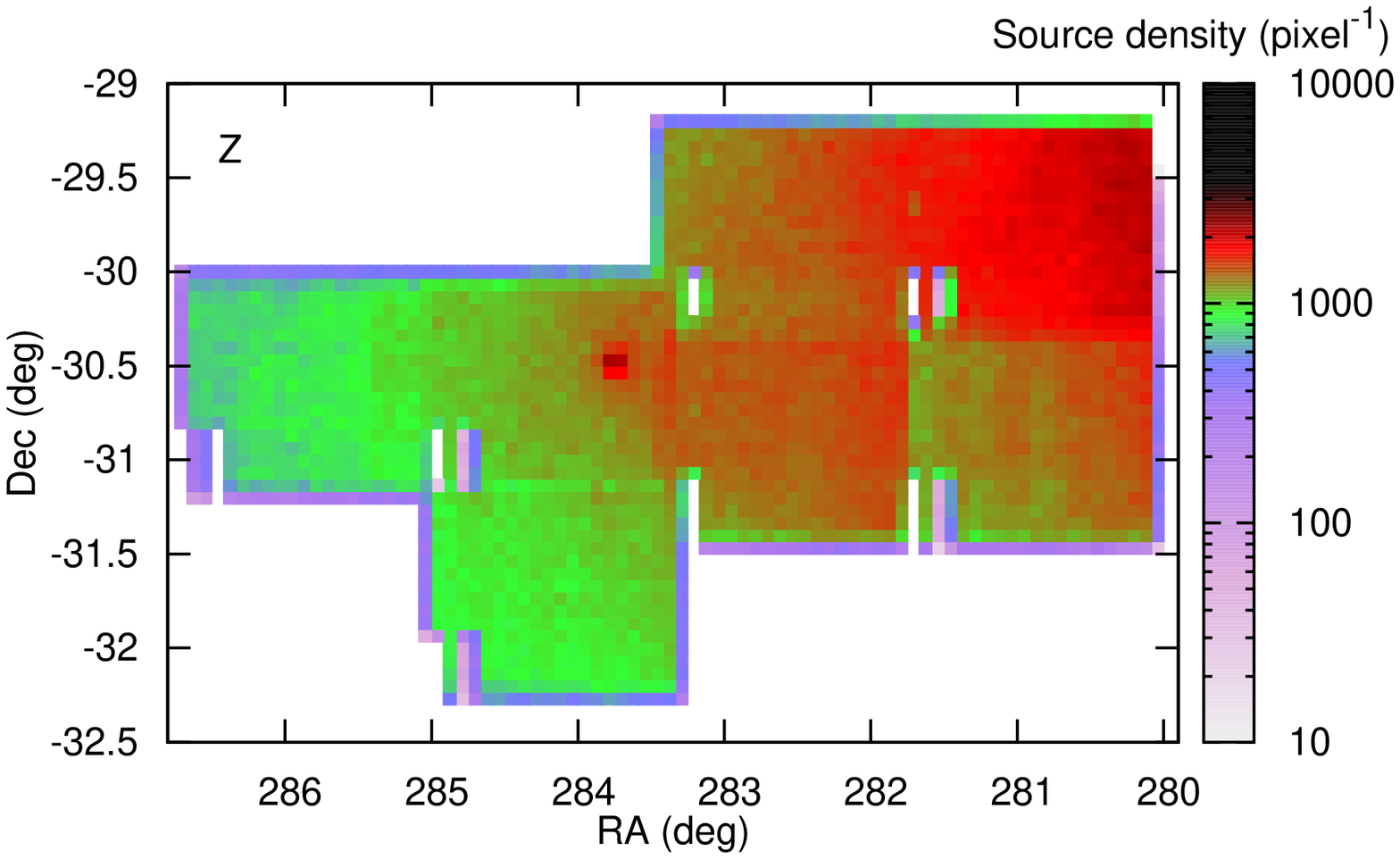}}
\vspace{-2mm}
\centerline{\includegraphics[height=0.70\textwidth,angle=-90]{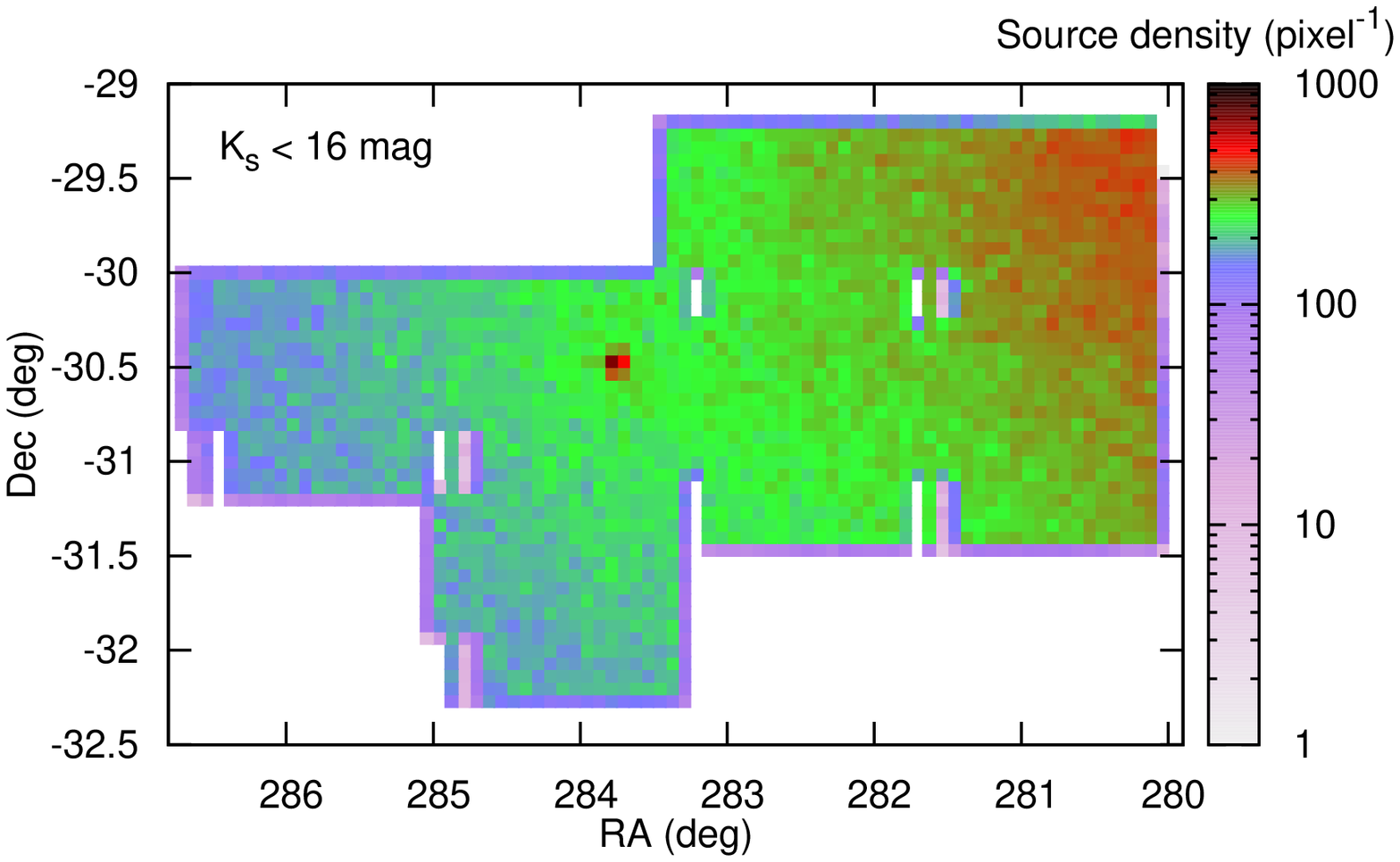}}
\caption{Distribution of sources identified in our final catalogue. Top panel: all sources, as detected in $Z$. Bottom panel: sources detected in $K_{\rm s}$ with $K_{\rm s} < 16$ mag. Note the tiling effect, caused by different adopted limiting magnitudes in each tile, is not present in the lower panel.}
\label{MapFig}
\end{figure*}

\begin{figure*}
\centerline{\includegraphics[height=0.93\textwidth,angle=-90]{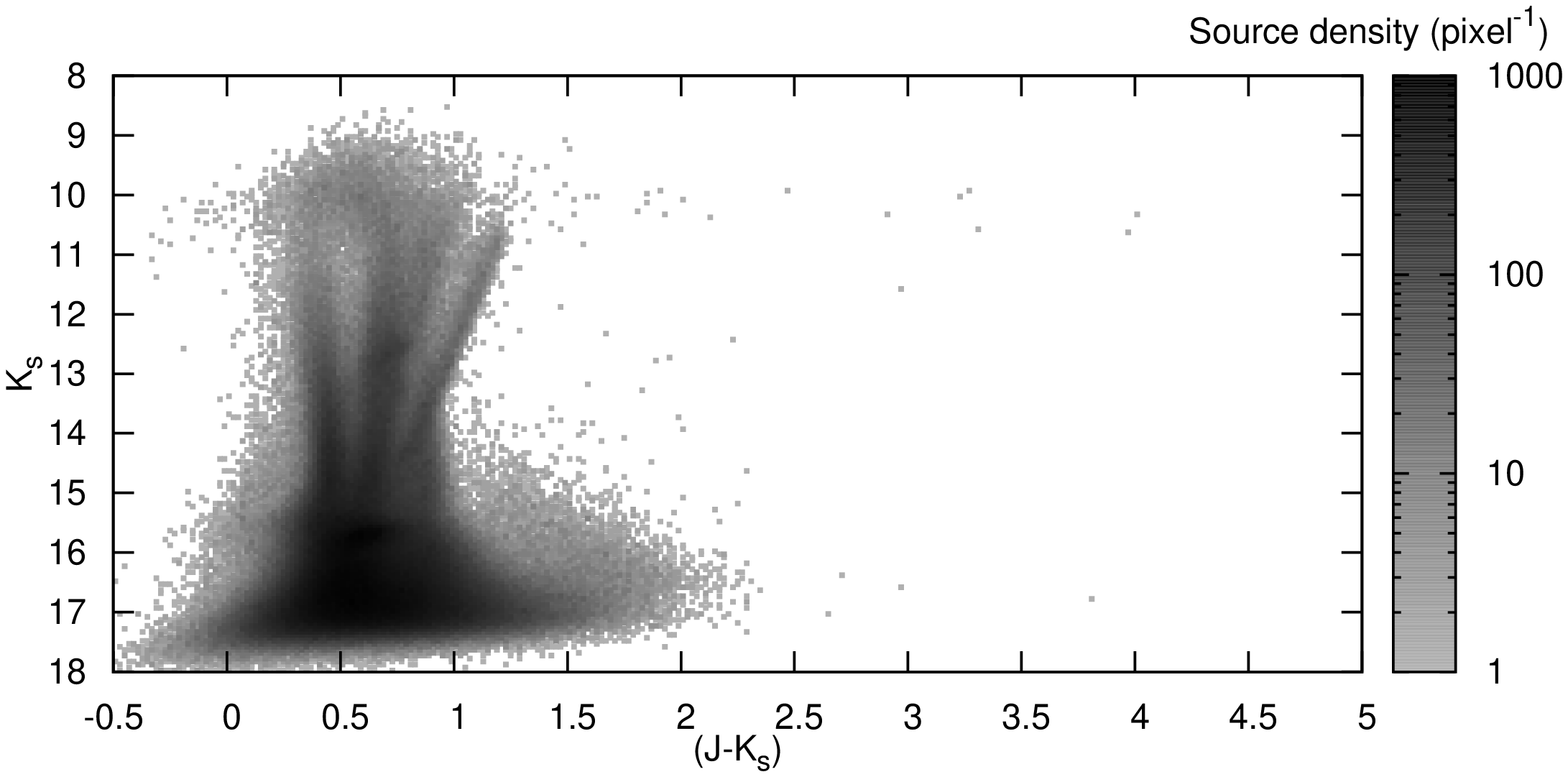}}
\vspace{-2mm}
\centerline{\includegraphics[height=0.93\textwidth,angle=-90]{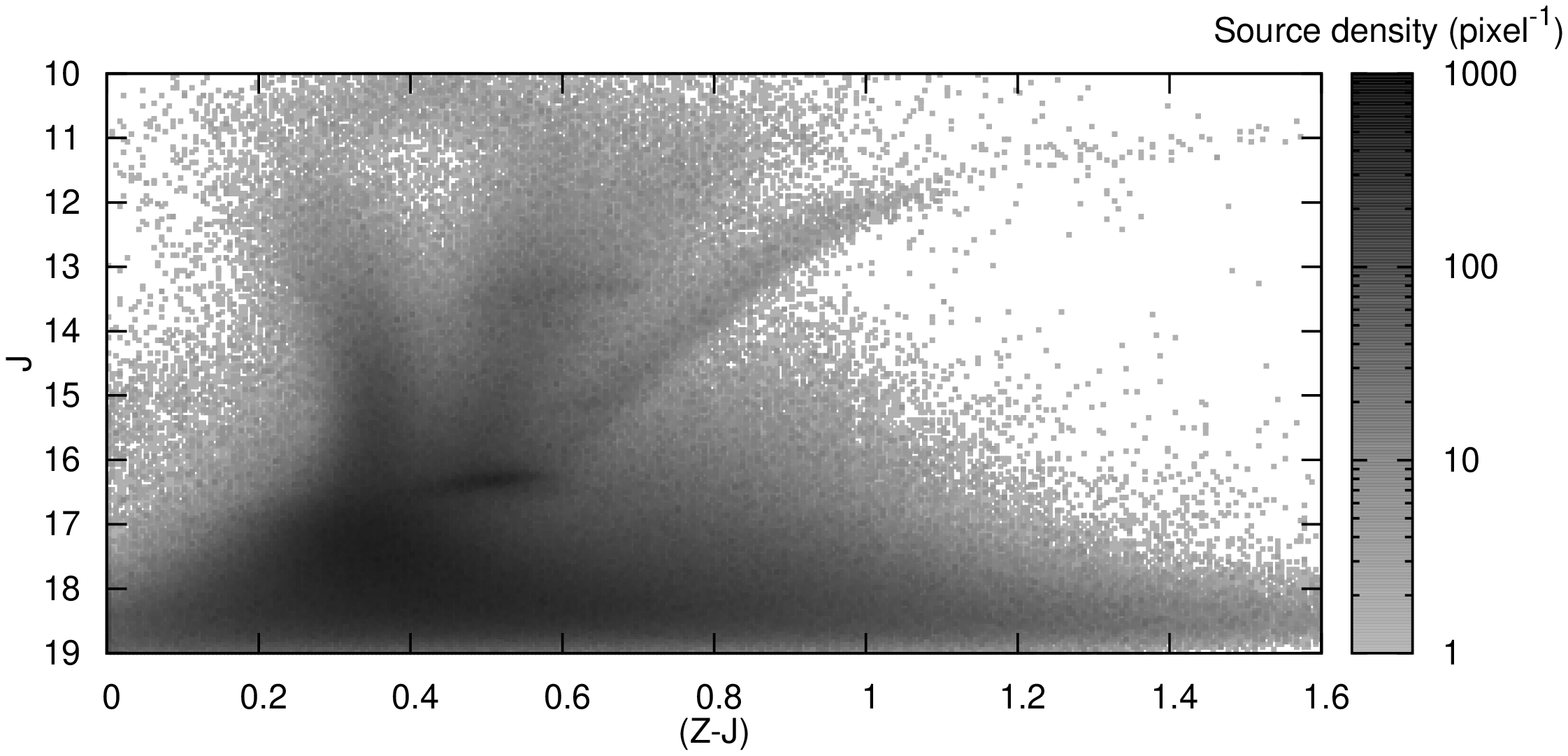}}
\caption{Colour-magnitude diagrams our from VISTA observations in the VISTA filter system.}
\label{CMDFig}
\end{figure*}

\begin{figure*}
\centerline{\includegraphics[height=0.47\textwidth,angle=-90]{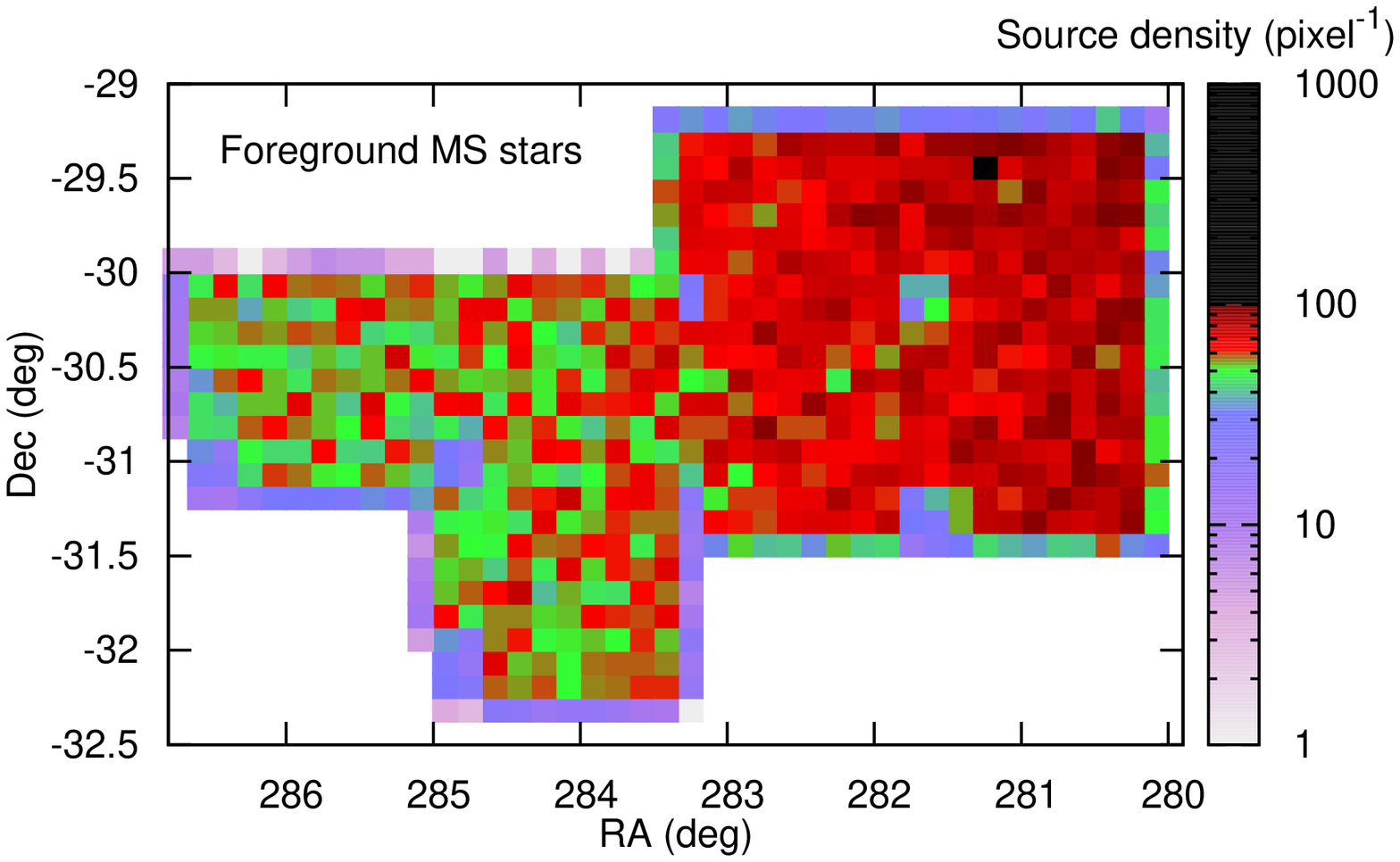}
            \includegraphics[height=0.47\textwidth,angle=-90]{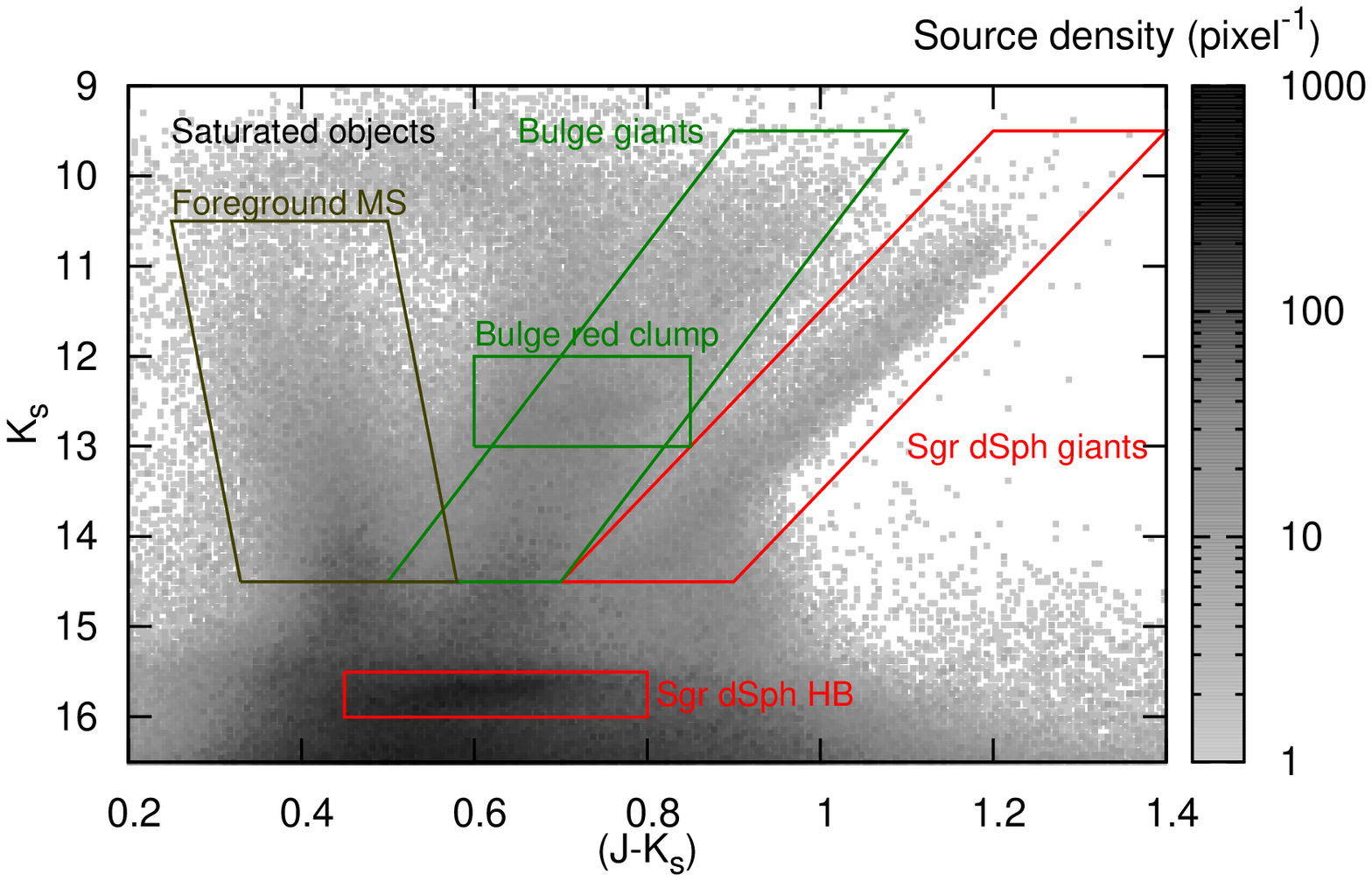}}
\vspace{-2mm}
\centerline{\includegraphics[height=0.47\textwidth,angle=-90]{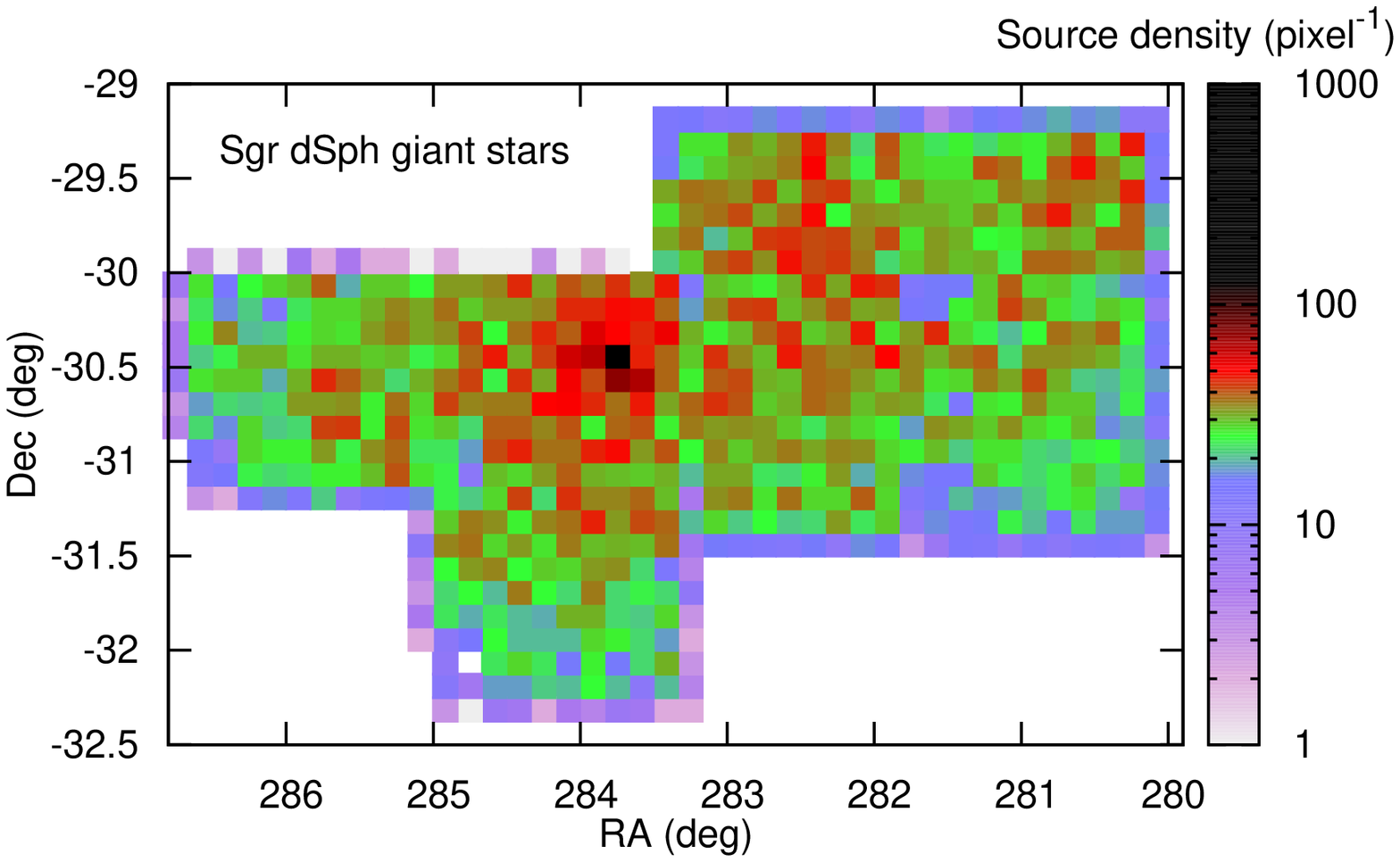}
            \includegraphics[height=0.47\textwidth,angle=-90]{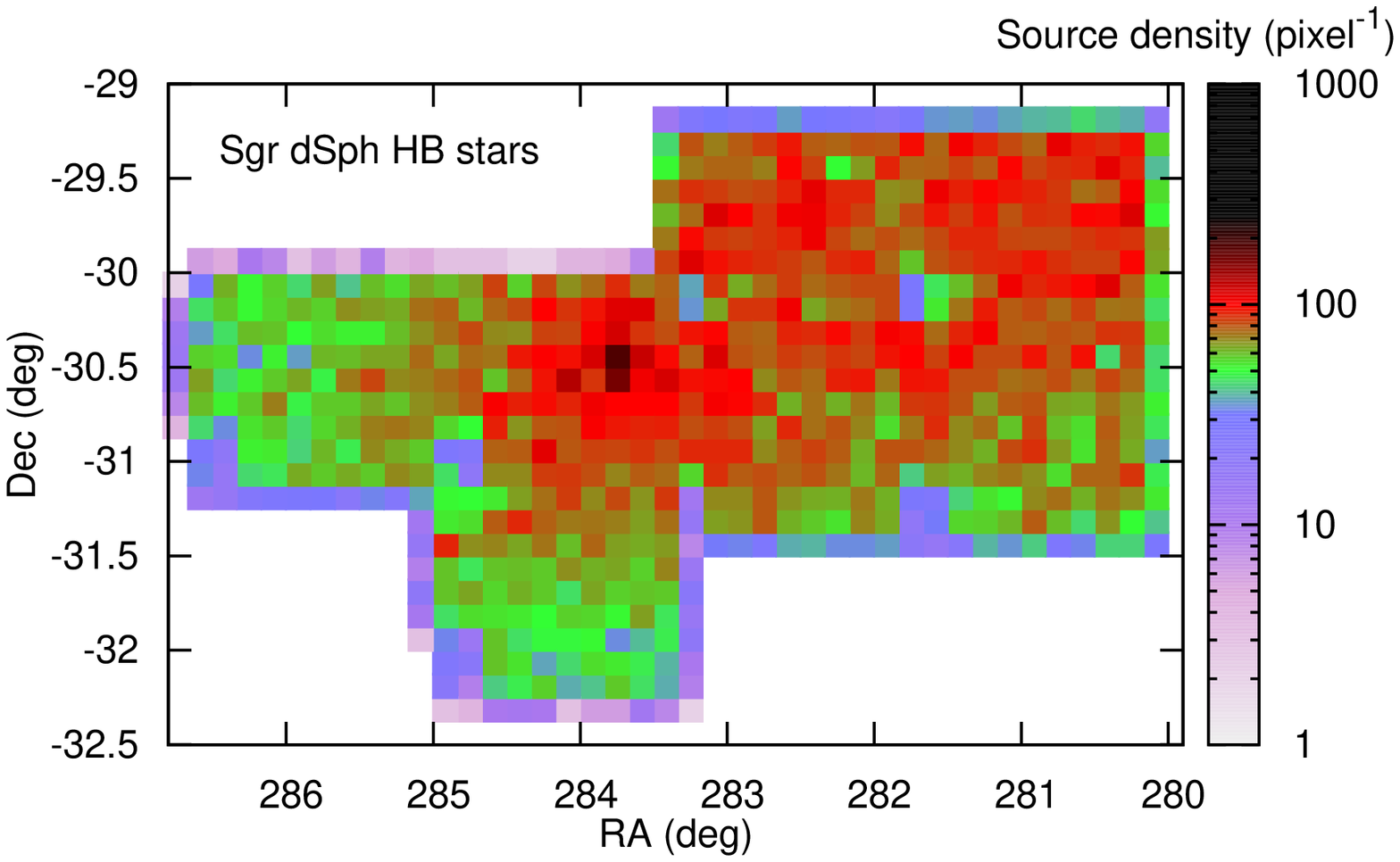}}
\vspace{-2mm}
\centerline{\includegraphics[height=0.47\textwidth,angle=-90]{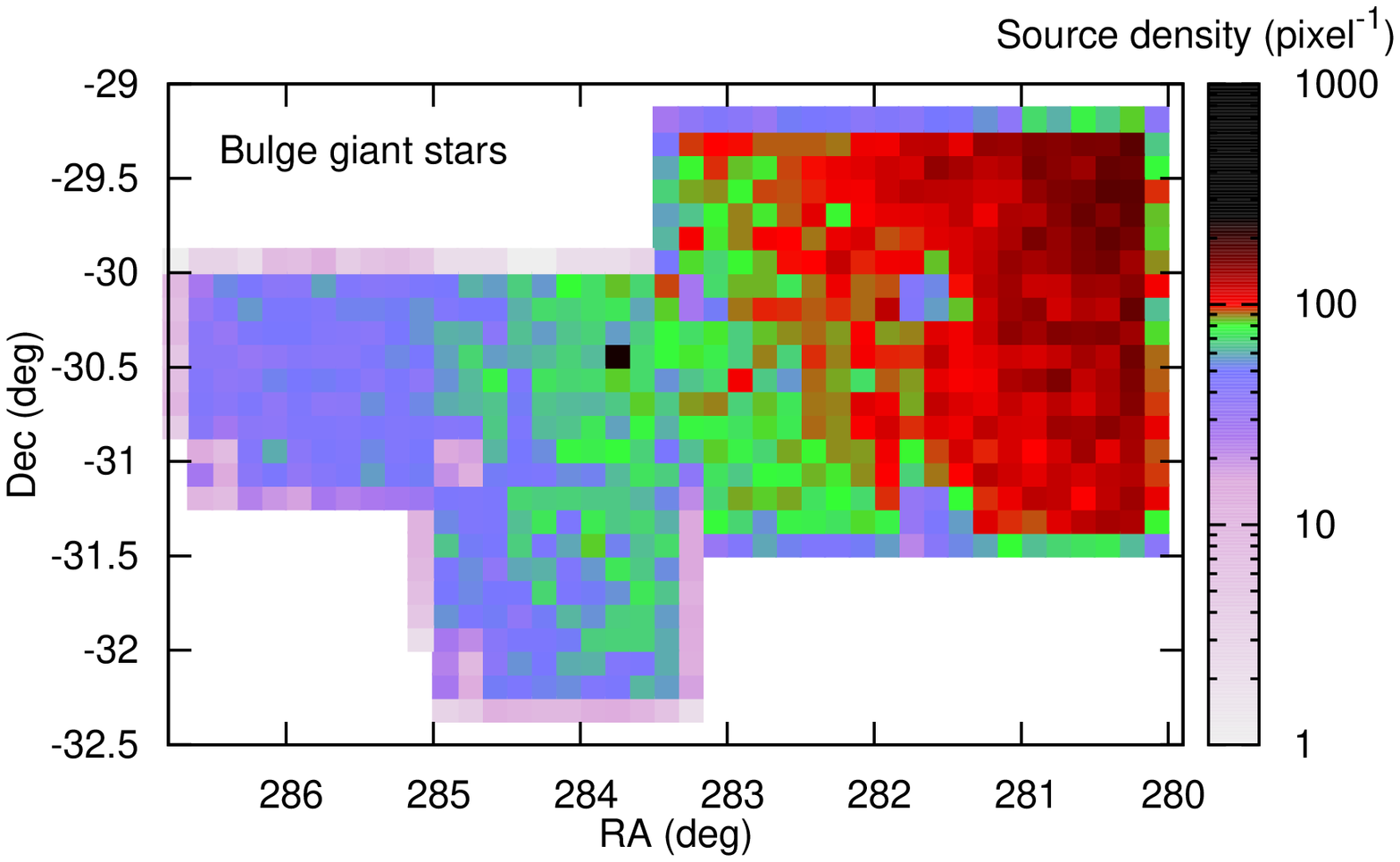}
            \includegraphics[height=0.47\textwidth,angle=-90]{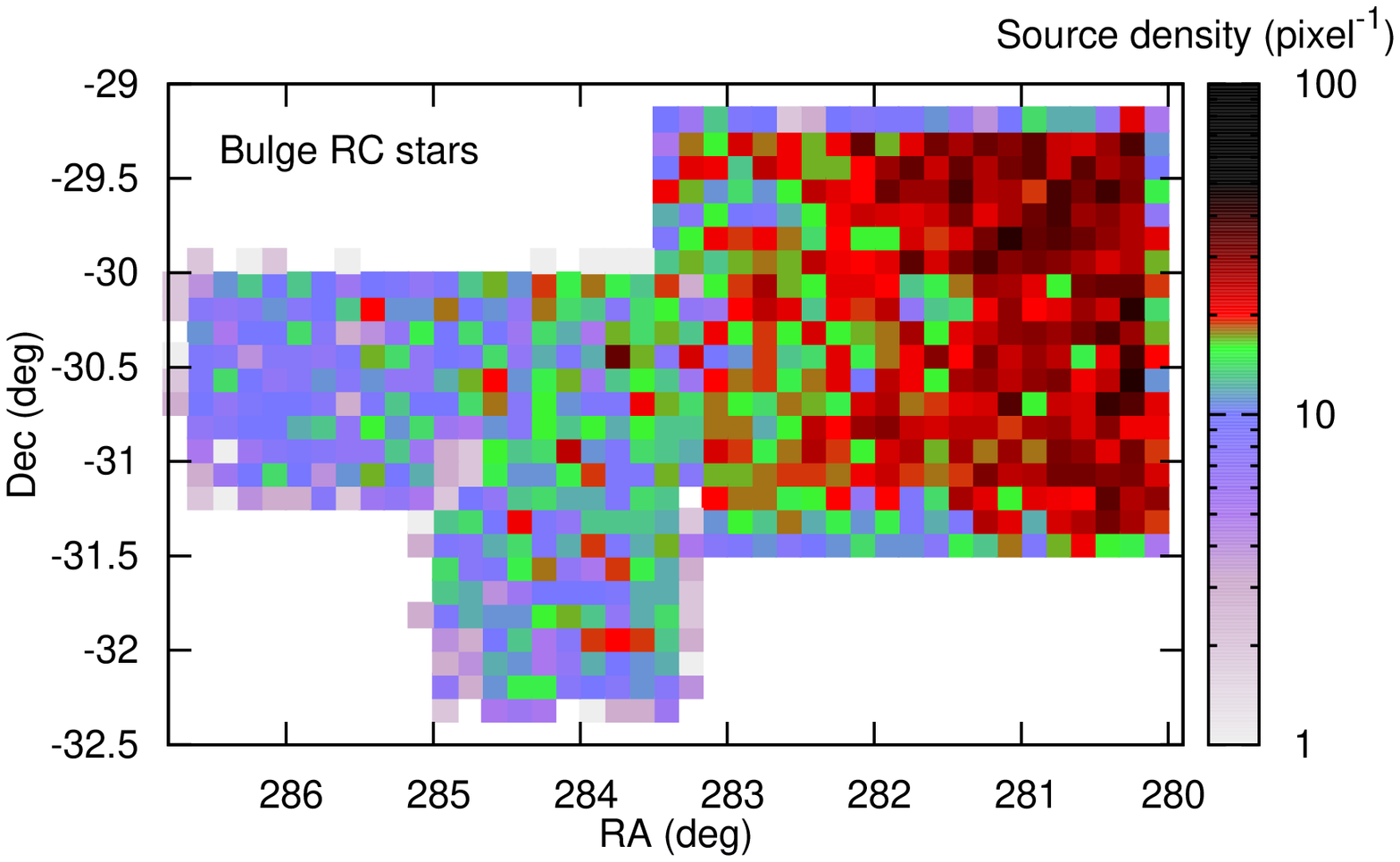}}
\vspace{-2mm}
\caption{Distribution of individual populations, as identified by regions of the colour--magnitude diagram in the top-right panel. The dense spot in the Sgr dSph panels marks M54, the galaxy's nominal centre. Sensitivity variations mean that a constant-density population will not be smooth over the images (see Figure \protect\ref{MapFig}). Abbreviations: MS --- main sequence; RC --- red clump; HB --- horizontal branch.}
\label{Map2Fig}
\end{figure*}

\begin{table*}
 \centering
 \begin{minipage}{160mm}
  \caption{Details of the identified sources, where the identifier contains the right ascension and declination in sexagesimal format. The continuation (below) contains the cross-identification ID and distance between source locations in major all-sky photometric catalogues. The full, contatenated table of all sources is available in the online version of the paper.}
\label{CatTable}
  \begin{tabular}{ccccccccccc}
  \hline\hline
   \multicolumn{1}{c}{ID}	&    \multicolumn{1}{c}{RA}	&    \multicolumn{1}{c}{Dec}
& \multicolumn{1}{c}{$Z$}	& \multicolumn{1}{c}{$\Delta Z$}	
& \multicolumn{1}{c}{$J$}	& \multicolumn{1}{c}{$\Delta J$}	
& \multicolumn{1}{c}{$K_s$}	& \multicolumn{1}{c}{$\Delta K_s$}\\
   \multicolumn{1}{c}{\ }	&    \multicolumn{1}{c}{Degrees}&    \multicolumn{1}{c}{Degrees}&
\multicolumn{1}{c}{(mag)} & \multicolumn{1}{c}{(mag)} & \multicolumn{1}{c}{(mag)} & \multicolumn{1}{c}{(mag)} & \multicolumn{1}{c}{(mag)} & \multicolumn{1}{c}{(mag)} \\
 \hline
VSgr18401401--3118205 & 280.0584 & --31.3057 & 12.357 & 0.030 & 12.377 & 0.288 & 11.813 & 0.372\\
VSgr18401620--3118259 & 280.0675 & --31.3072 & 10.925 & 0.033 & 10.895 & 0.302 & 10.211 & 0.359\\
VSgr18401855--3121500 & 280.0773 & --31.3639 &  9.169 & 0.171 & 09.542 & 0.224 &\nodata &\nodata\\
\nodata & \nodata & \nodata & \nodata & \nodata & \nodata & \nodata & \nodata & \nodata \\
\hline
\end{tabular}
  \begin{tabular}{ccccccccccc}
   \multicolumn{1}{c}{ID}	
& \multicolumn{2}{c}{UCAC4} & \multicolumn{2}{c}{2MASS} & \multicolumn{2}{c}{WISE}\\
   \multicolumn{1}{c}{\ }	& \multicolumn{1}{c}{ID} & \multicolumn{1}{c}{($^{\prime\prime}$)} & \multicolumn{1}{c}{ID} & \multicolumn{1}{c}{($^{\prime\prime}$)} & \multicolumn{1}{c}{ID} & \multicolumn{1}{c}{($^{\prime\prime}$)}\\
 \hline
VSgr18401401--3118205 & 294-200130 & 0.187 & 18401400--3118206 & 0.166 & J184014.00--311820.6 & 0.243\\
VSgr18401620--3118259 & 294-200153 & 0.191 & 18401618--3118260 & 0.212 & J184016.18--311826.1 & 0.202\\
VSgr18401855--3121500 & 294-200171 & 0.183 & 18401853--3121501 & 0.250 & J184018.53--312150.1 & 0.168\\
\nodata & \nodata & \nodata & \nodata & \nodata & \nodata & \nodata \\
\hline
\end{tabular}
\end{minipage}
\end{table*}

The final catalogue (Table \ref{CatTable}) contains 2\,921\,920 VISTA objects, of which 1\,047\,713 have detections in all three bands (the sharp decrease is due to the much lower efficiency in detecting intrinsically fainter sources given higher sky background at longer wavelengths). Of the VISTA objects, 433\,245 have identified 2MASS counterparts, 48\,028 have UCAC4 counterparts, and 222\,191 have identified \emph{WISE} counterparts.

Figure \ref{MapFig} shows the projected density of VISTA sources. Source density can be seen to increase to the north-west (upper right) where the Galactic Bulge contributes an increasing number of stars. Gaps can be seen where coverage is decreased or absent due to removal of a poorly-functioning detector on the VIRCAM CCD. The high-density region at $\alpha = 283.75$, $\delta = -30.5$ is M54, which marks the galaxy's core.

Figure \ref{CMDFig} presents the associated colour--magnitude diagrams, with the distribution of component populations plotted in Figure \ref{Map2Fig}. Considerable contamination is evident along the line of sight from both the Galactic Plane and Galactic Bulge.

The foreground main sequence is mostly confined to $(J-K_s) < 0.6$ mag. Its stars are distributed more uniformly across the surveyed area than the other populations, but shows concentrations towards high Galactic latitudes, which lie towards the top-right of the mapped panels in Figure \ref{Map2Fig}.

The Sgr dSph giant branches are well defined and easily separable from the other populations at the bright end, near the RGB tip. The asymptotic giant branch (AGB) continues above the RGB and curves off to the red side of the diagrams, with the curve being quite sharp when plotted as $K_s$ versus $(J-K_s)$. All these red (($J-K_s$) $\gtrsim$ 1.3) AGB stars are thought to be carbon-rich \citep{MWZ+12}. The stars immediately above the RGB tip contain a mixture of oxygen- and carbon-rich AGB stars within the Sgr dSph. The base of the AGB can be faintly seen at around $(J-K_s) = 0.8$ and $K_s = 14.4$ mag, also at $(Z-J) = 0.6$ and $J = 15.2$ mag. The Sgr dSph giant stars are concentrated around M54 (Figure \ref{Map2Fig}, centre-left panel), but also reveal the greater extent of the galaxy further from the cluster. Structure can be seen on various scales at roughly twice the level given by Poisson noise. Unfortunately, the differing sensitivity limits and irregular footprint of our observations means we are limited in the quantitive detail we can place on the outer regions of the Sgr dSph galaxy. Additionally, the low contrast of these features means determining the shape, extent and origin of that structure is not reliably possible. Altering the lower-magnitude boundary of stars included in this map does not have any appreciable impact on the existence or otherwise of any structures.

The Bulge giant branch dominates the central and upper parts of the colour--magnitude diagram. Its stars clearly concentrate towards the Galactic Centre, with a density variation of a factor of around ten over the mapped area. In colour--magnitude space, it disappears near $K_{\rm s}$ = 16 mag, where the three populations converge. Few Bulge stars are seen at southerly Galactic latitudes. Here, stars identified as Bulge giants are likely giants in the Galactic Plane: even though the Plane is typically closer (therefore brighter) than then Bulge, there is still sufficient overlap that the Plane contaminates the Bulge giant population. The full colour--magnitude diagrams (Figure \ref{CMDFig}; see also later, Figure \ref{BiasFig}) extend down to the Bulge main-sequence turn off, which is just visible at the faint limit of the $K_s$-band data, but is more obvious near $J$ = 18 mag in the shorter-wavelength data.

Two final density enhancements are present in Figure \ref{CMDFig}, labelled as the Bulge red clump and the Sgr dSph horizontal branch in Figure \ref{Map2Fig}. In reality, these represent the combination of two different phenomena. Evolutionarily, the first is the RGB bump, caused by chemical discontinuities in the hydrogen-burning shell as a result of first dredge-up. The second is the horizontal branch (HB), or start of the helium-burning period of evolution. This is termed the ``red clump'' in metal-rich and higher-mass populations where no significant horizontal extent to the branch is seen.

In the Bulge, we mainly see the RGB red bump (see Section \ref{IsoBulgeSect}), as the red clump stars are more vertically spread in the colour--magnitude diagrams. In the Sgr dSph, we primarily see the HB stars: the RGB bump is just visible below the HB in the $J$ vs.\ ($Z-J$) colour--magnitude diagram. The extent of the Sgr dSph HB is rather less in $(Z-J)$ than in $(J-K_s)$. This extension in $(J-K_s)$ is unlikely to be real and probably results from decreased sensitivity in the longer-wavelength bands. Using colour-cuts of $0.44 \leq (Z-J) \leq 0.56$ and $16.15 \leq J \leq 16.60$ mag, we have produced the map of Sgr dSph HB stars in the centre-right panel of Figure \ref{Map2Fig}. Although it has more sources, this map shows a strong concentration of sources around M54. However, it also suffers slightly more from contamination, leading to an erronous increase in sources in the top right.



\section{Isochrone fitting}
\label{IsoSect}

\subsection{Separating the populations in the colour--magnitude diagram}
\label{IsoBSect}

\begin{figure*}
\centerline{\includegraphics[height=0.80\textwidth,angle=-90]{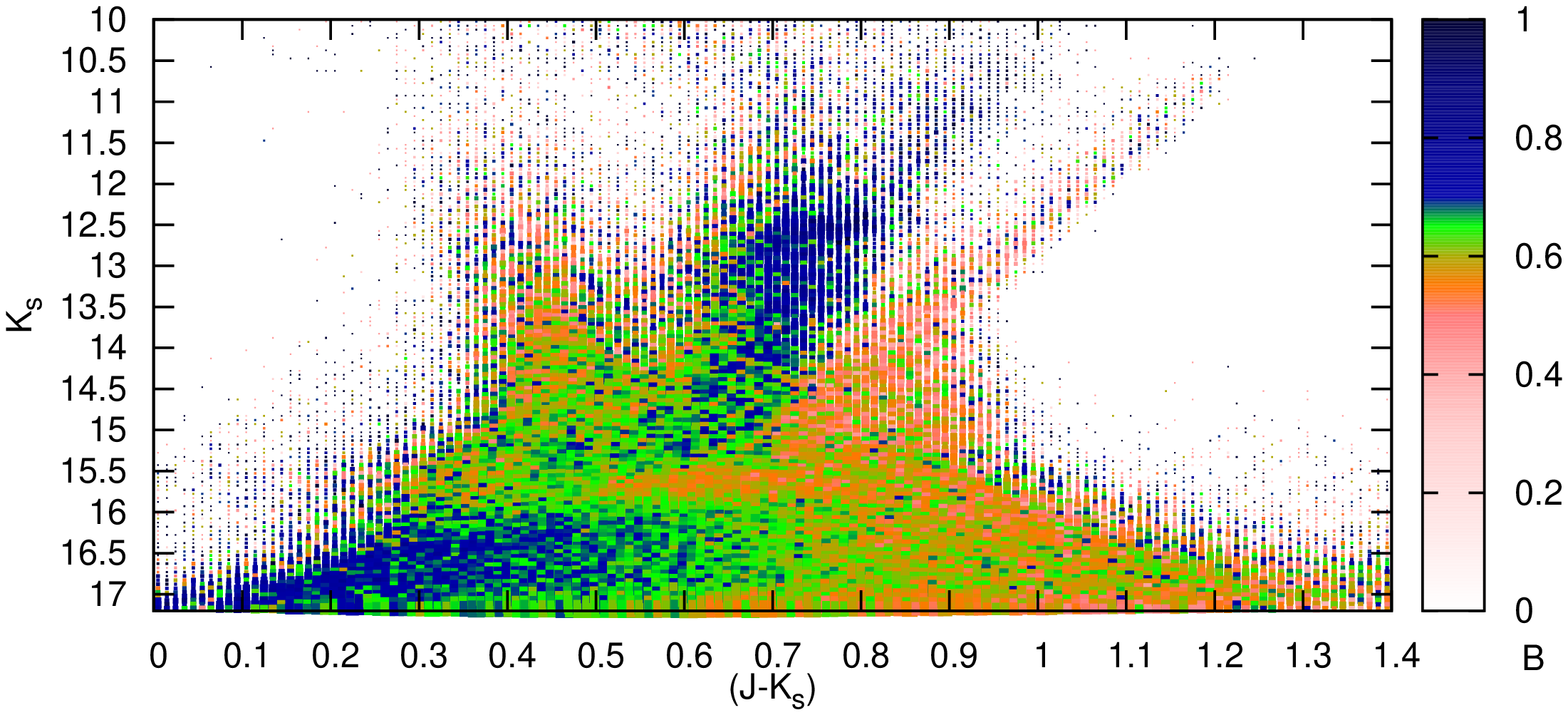}}
\centerline{\includegraphics[height=0.49\textwidth,angle=-90]{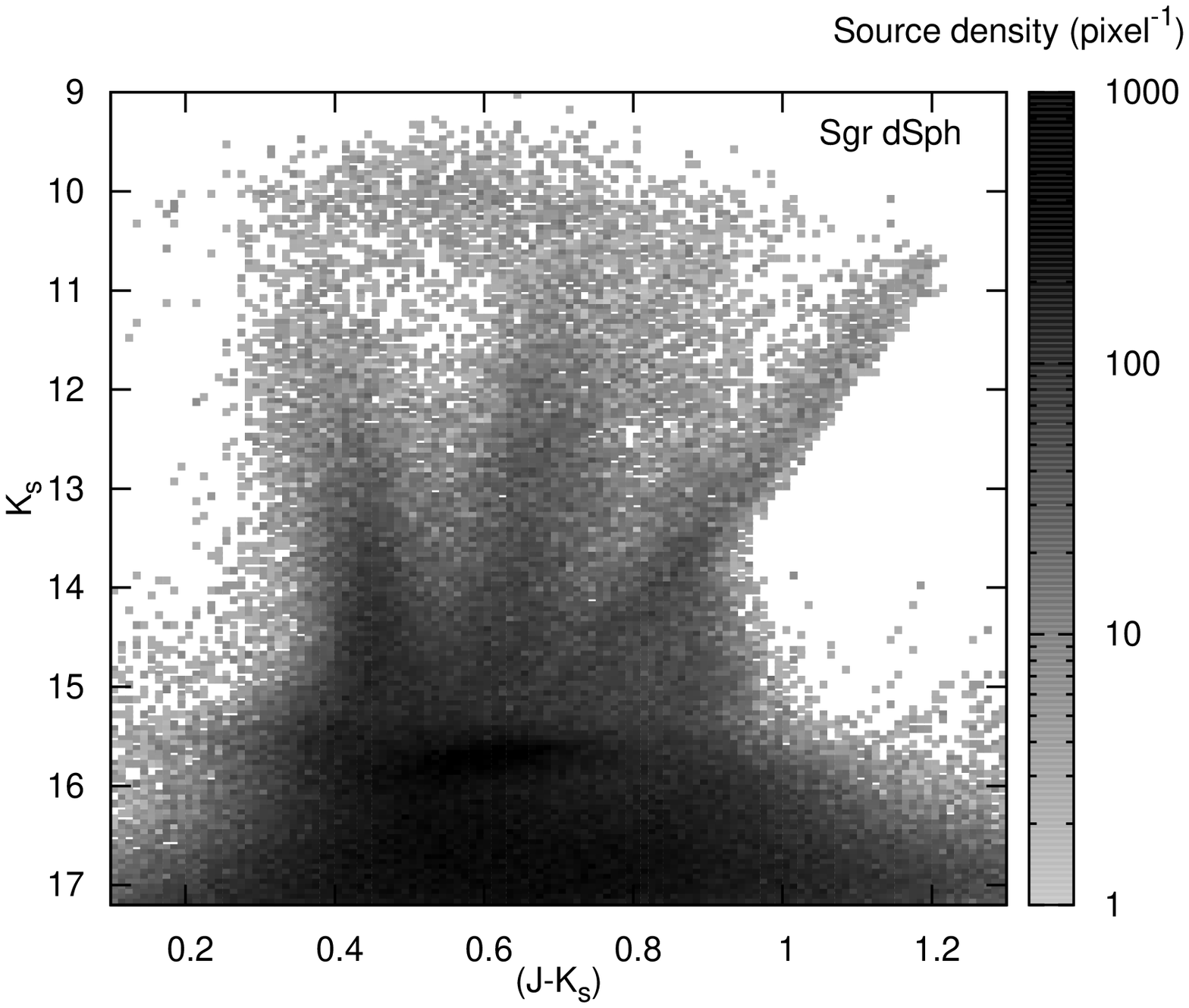}
            \includegraphics[height=0.49\textwidth,angle=-90]{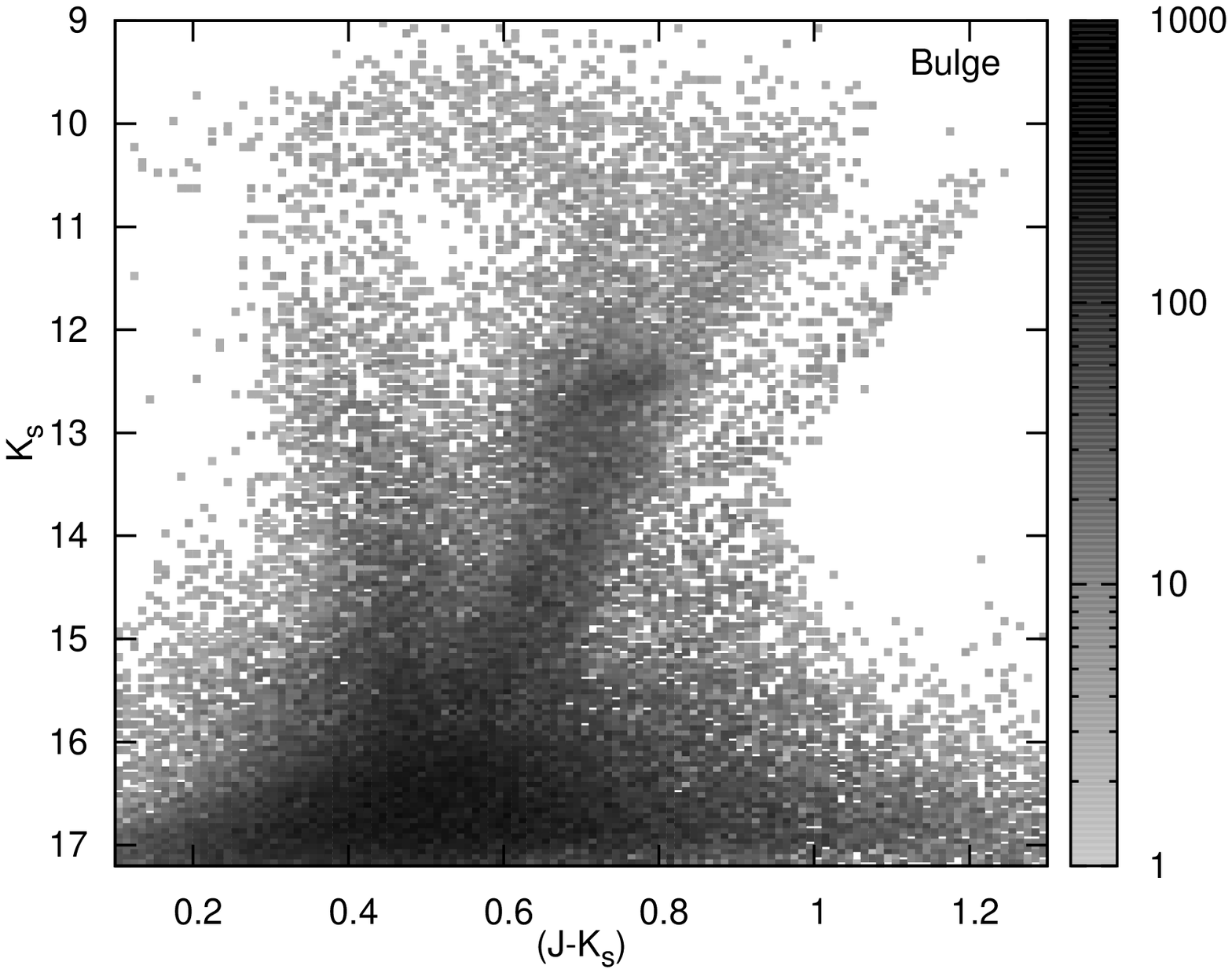}}
\centerline{\includegraphics[height=0.49\textwidth,angle=-90]{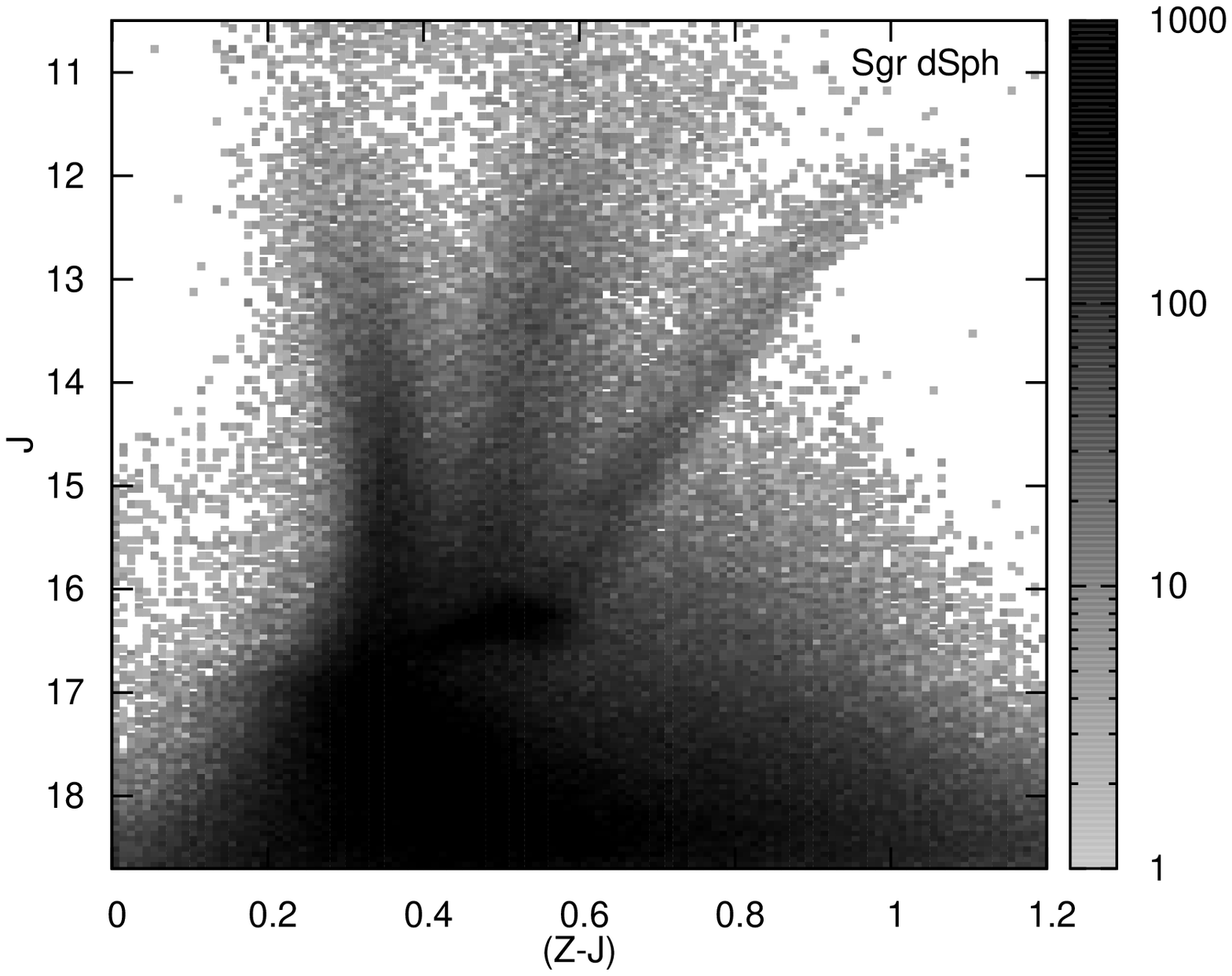}
            \includegraphics[height=0.49\textwidth,angle=-90]{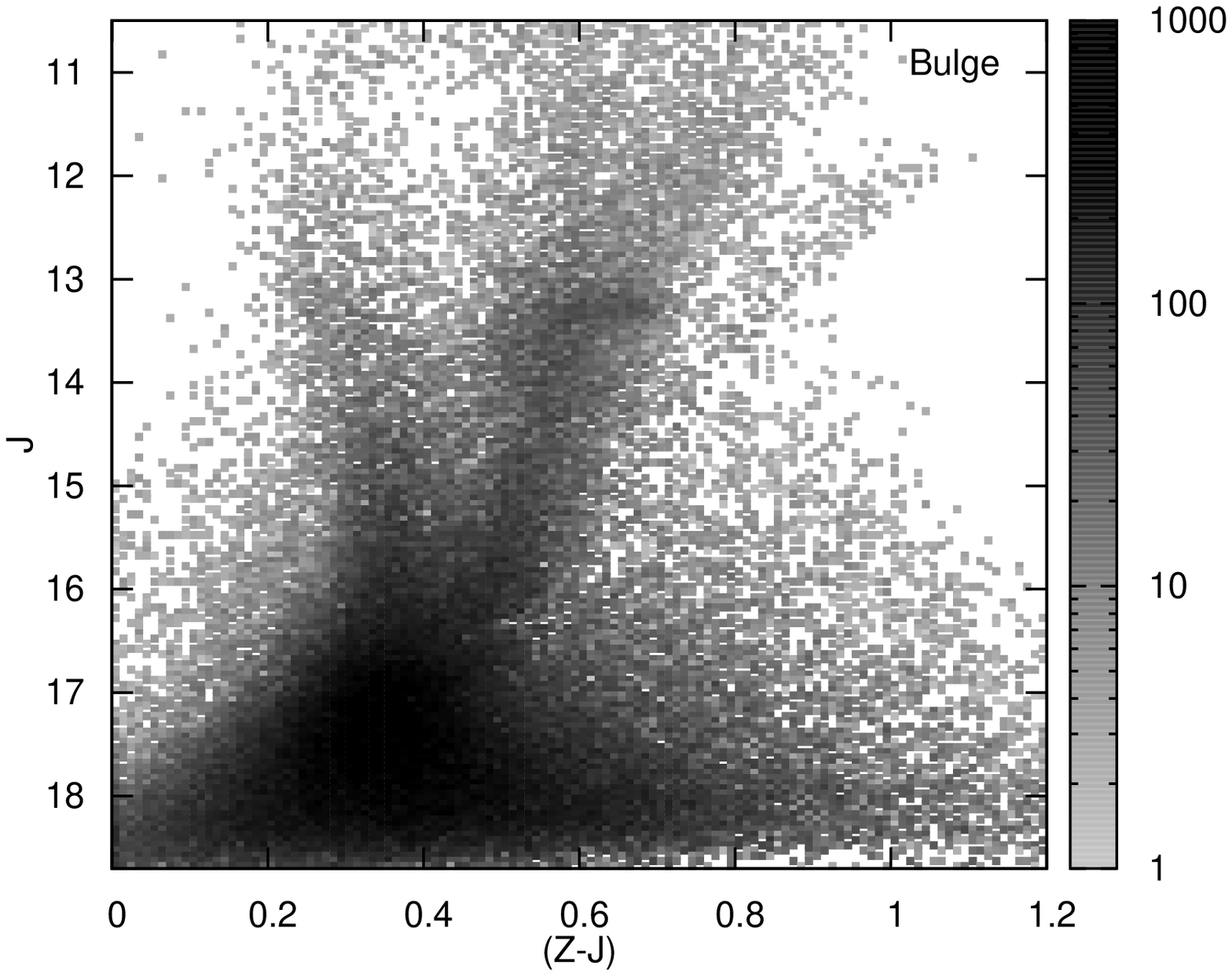}}
\caption{Top panel: colour--magnitude diagrams showing separation of stars into stars with Sgr-dSph-like populations (colour scale, $B \sim 0.5$) and Bulge-like populations ($B \sim 1$). Point area scales with the number of stars in each bin. Middle panels: statistically extracted populations of Sgr dSph and Galactic Bulge stars based on $B$ values from the top panel. Foreground stars remain in both extracted populations, but are biased towards appearing in the Sgr dSph population. Bottom panels: as middle panels, for $(Z-J)$ versus $J$.}
\label{BiasFig}
\end{figure*}

In order to fit accurate isochrones to the observed colour--magnitude diagrams, we have endeavored to separate the Bulge and Sgr dSph populations from each other, by bisecting our sample in Galactic latitude around M54. The population at southerly Galactic latitudes should contain mostly Sgr dSph stars, the population at northerly Galactic latitudes (closer to the Galactic Plane) will be dominated increasingly by the Galactic Bulge.

As we have no field of solely Sgr dSph nor Bulge stars to use as a template, we can only determine the bias of a particular combination of colour and magnitude being present in the southern or northern fields (therefore being in the Sgr dSph or Bulge). We define the following statistic:
\begin{equation}
B(J-K_{\rm s}, K_{\rm s}) = \frac{\rho_N}{\rho_N + \rho_S},
\label{BiasEq}
\end{equation}
where $\rho_N$ and $\rho_S$ are the density of sources per unit area of sky in the (Galactic co-ordinate) northern and southern regions. This defines the amount of northerly bias present in the sample.

The total source density is strongly biased towards regions near the Galactic Equator, with the northern region containing 63\% of the stars, and the southern half 37\%. The Sgr dSph has its major axis in the direction of Galactic latitude, and M54 is its nominal centre \citep{IGI95}. If its population is symmetrically distributed across in both halves, its population should have values of $B$ of 0.50. The Bulge should be predominantly in the northern regions, thus it should have $B = 0.63$ to 1. The Galactic Disc should show some preference for the Bulge populations due to a longer line of sight, but also contains a local population of stars which are more uniformly distributed. It should therefore have values of $B$ between those of the Sgr dSph and the Bulge.

Figure \ref{BiasFig} illustrates the north-south bias for our sample. The Sgr dSph giant branch is clearly defined in this plot, visible down to the HB, which juts blueward of the giant branch near $K_{\rm s} = 16$ mag. The Sgr dSph population becomes much harder to identify below the HB as it merges into the base of the Bulge RGB. The upper bulge RGB is also well-defined, but merges into the Galactic Disc RGB on its blueward edge. The Galactic MS, Bulge RGB and Sgr dSph HB all co-incide, making it difficult to identify the blue edge of the Sgr dSph HB. A uniformly distributed extension of stars near $(J-K_{\rm s}) = 0.25$, $K_{\rm s} = 16$ mag may be a metal-poor extension of the Sgr dSph population, or it may truncate near $(J-K_{\rm s}) = 0.50$ mag. Stars in M54, with metallicities much below the Sgr dSph average, lie on the redward side of the Bulge RGB.

Any test for asymmetry in the Sgr dSph population is hampered by the fact that the survey area is not symmetrical about the Galactic north--south axis. With this concession in mind, we can only perform a limited test for asymmetry between the two halves. Taking stringent limits on the stars within the Sgr dSph giant branch, namely $21.5 < K_{\rm s} + 10(J-K{\rm s}) < 23.5$ mags with $(J-K_{\rm s}) > 0.95$ mags and $K_{\rm s} > 10.5$ mags, we find 2759 stars with an average $B = 0.519 \pm 0.019$, implying no measurable asymmetry in the Sgr dSph galaxy population. In practice, since the centroid of the northern half of the observations is considerably further from M54 than the southern centroid, we should expect it to have a slightly lower density, thus $B$ should be slightly less than 0.5. Thus, it seems likely that there are more stars in the Sgr dSph at higher Galactic latitudes than there are at more-negative Galactic latitudes, implying M54 is not the true centroid of the Sgr dSph.

Although we have no control field to subtract, Figure \ref{BiasFig} shows that sections of the Bulge giant branch reach values of $B$ which approach unity. A variety of ranges can be placed onto the colour--magnitude diagram where $B$ is maximised. This effectively sets a lower limit to the fraction of Bulge stars in the northern section. The 1-$\sigma$ lower limit is maximised near the region $12.35 \leq K_{\rm s} \leq 12.55$ and $0.79 \leq (J-K_{\rm s}) \leq 0.81$, containing 214 stars, where $B$ is 0.921 $\pm$ 0.068. This sets a 1-$\sigma$ limit whereby at least 85.2\% of stars with those parameters (and by implication all Bulge stars) are in the northern region.

To first order, we can use this to extract both the Sgr dSph and Bulge populations from our sample. The confusing issue of the Galactic Plane means that some of this population will be present in both extracted populations. Assuming 92.1\% of the Bulge and 50\% of the Sgr dSph stars are in the northern half, given a total number of stars $N_{\rm total}(J-K_{\rm s}, K_{\rm s})$ in a particular colour--magnitude bin, the number of stars in that bin belonging to each population should be given by:
\begin{eqnarray}
N_{\rm Sgr}(J\!-\!K_{\rm s}, K_{\rm s}) \!\!\!\!&=&\!\!\!\! N_{\rm total} \left( 1 - \frac{B - 0.500}{0.921 - 0.500}\right) , \nonumber \\
N_{\rm Bulge}(J\!-\!K_{\rm s}, K_{\rm s})     \!\!\!\!&=&\!\!\!\! N_{\rm total} \frac{B - 0.500}{0.921 - 0.500} .
\label{Bias2Eq}
\end{eqnarray}

The lower panels of Figure \ref{BiasFig} show the results of this statistical extraction. Bulge stars do appear to have been cleanly extracted from the rest of the data, while the foreground stars have gravitated towards the Sgr dSph population as a result of having similar values of $B$. In each case, the faintest stars have fallen within the Sgr dSph population, due to the fainter limiting magnitude in these less-crowded tiles. An attempt was made to separate the foreground stars from the Sgr dSph stars using similar techniques. This test used the average projected radius from M54 as the diagnostic, instead of Galactic latitude. However, it proved not to be very decisive, given the limited spatial coverage of our observations compared to the size of the Sgr dSph.

\subsection{The Sgr dSph}
\label{IsoSgrSect}

\subsubsection{Comparison to literature data}
\label{IsoSgr1Sect}

\begin{figure*}
\centerline{\includegraphics[height=0.97\textwidth,angle=-90]{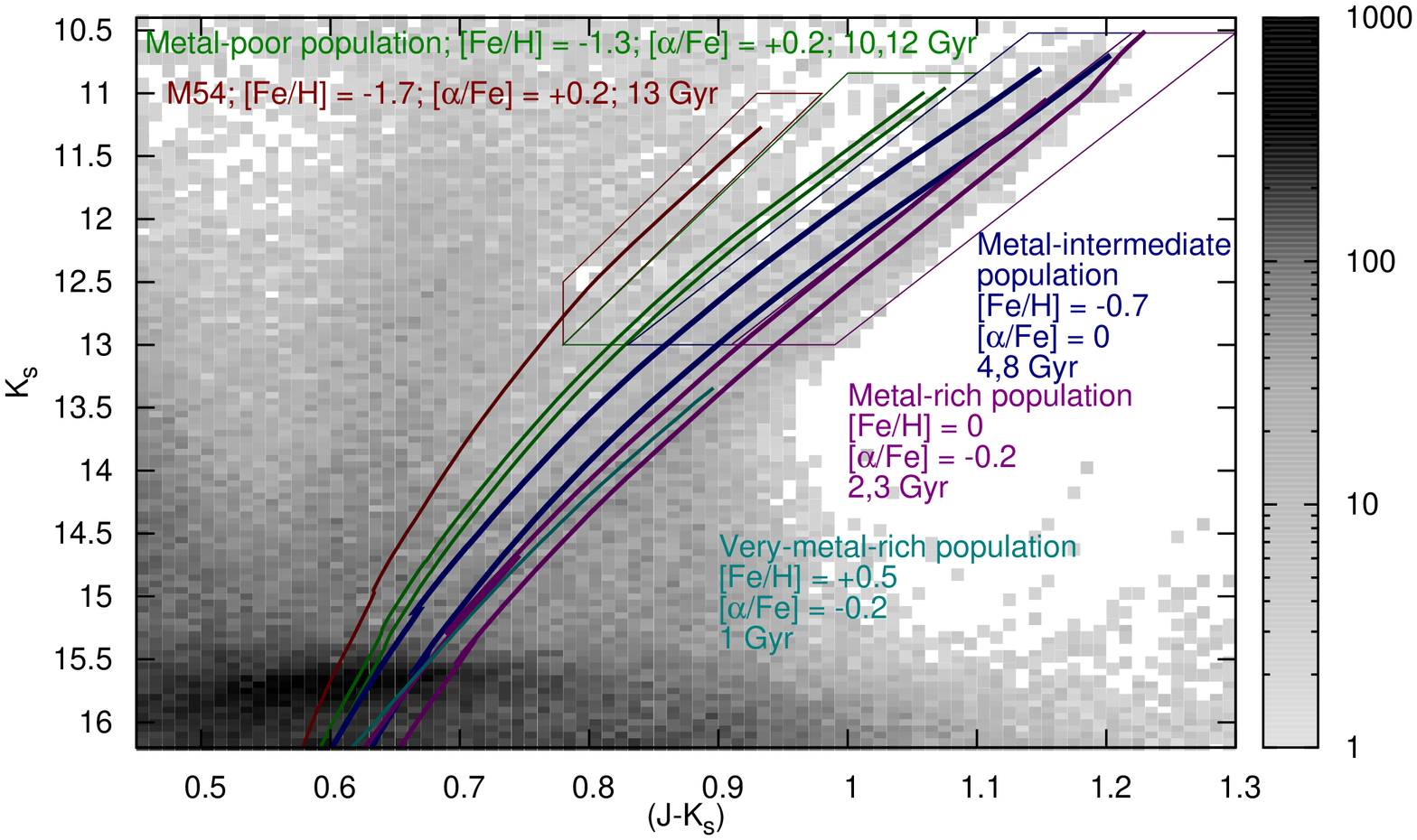}}
\centerline{\includegraphics[height=0.97\textwidth,angle=-90]{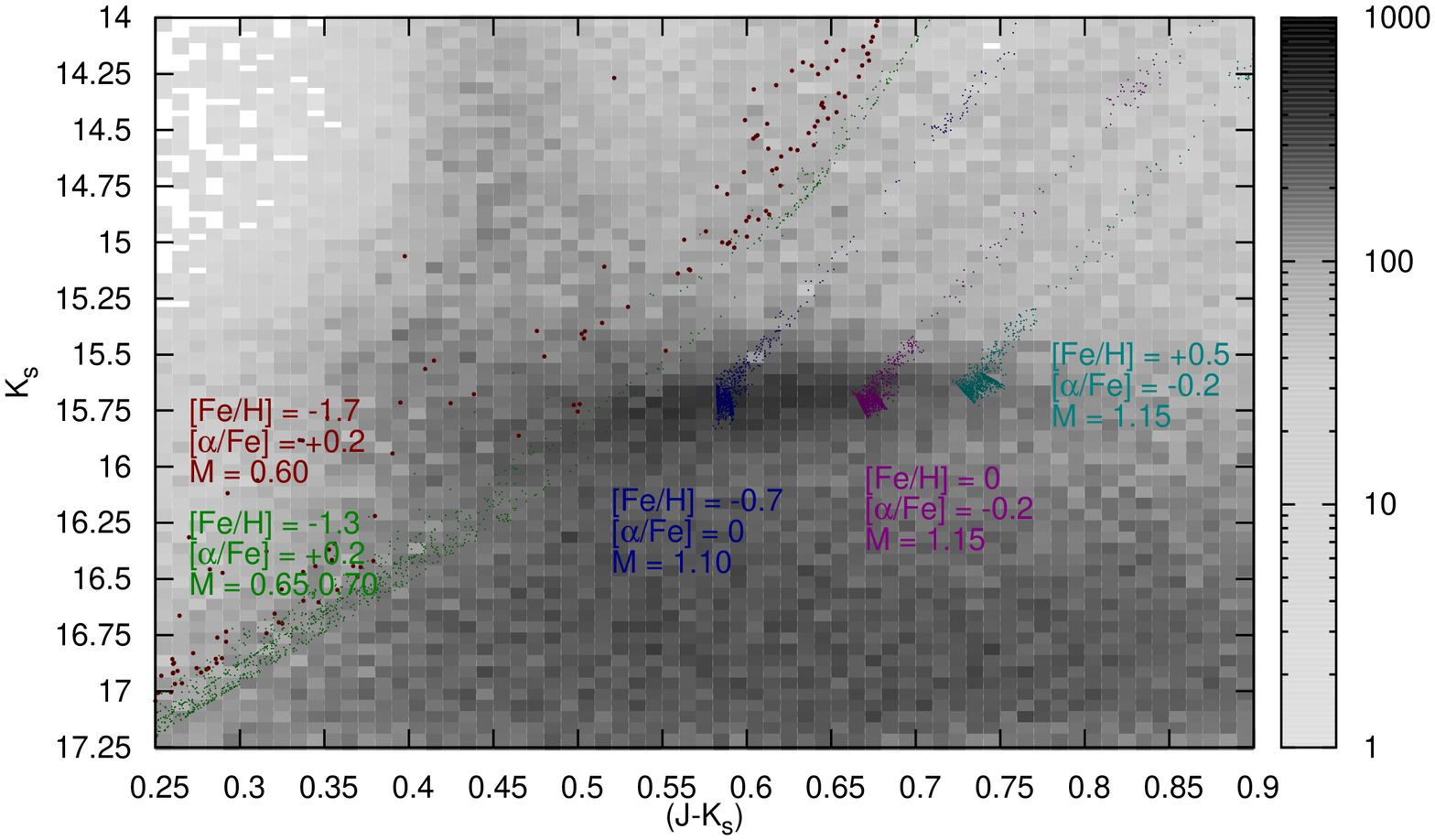}}
\caption{Dartmouth stellar isochrones and horizontal branch models fit to the statistically-extracted Sgr dSph population from Figure \protect\ref{BiasFig}. The five known populations from SDM+07 are presented, both from left to right and order of increasing metallicity, by red, green, blue, magneta and cyan lines and points. Details on the isochrones are given in the text. A colour version of the figure is available in the online edition of this work.}
\label{SgrIsoFig}
\end{figure*}

The five-population summary present by SDM+07 represents the cumulation of a number of previously published studies, along with their own work. These estimates can be broadly summarised as follows:

{\bf Distance:} early estimates using red giants have put the Sgr dSph consistently at distances from 25.2 $\pm$ 2.8 kpc \citep{MUS+95} to 26.3 $\pm$ 1.8 kpc \citep{MBFP04}. More recent work using RR Lyrae has sugggested a slightly closer distance (24.8 $\pm$ 0.8 kpc; \citealt{KC09}), while \emph{Hubble Space Telescope} data has been re-analysed by \citet{SML+11} to give a significantly further distance (28--30 kpc, inclined towards our line of sight).

{\bf Metallicity distribution:} the metallicity distribution has been determined by fitting globular cluster loci or isochrones to colour--magnitude diagrams by several authors (\citealt{SL95,MUS+95,LS97,BFB99a,BFB99b,LS00,BNC+06}; SDM+07; \citealt{GSZ+10,SML+11}). The earlier works among these sampled few stars in small regions around the galaxy, and did not accurately consider stars' $\alpha$-element enhancement. Thus while they found a distribution of metallicities in the galaxy, they did not always agree on the bulk metallicity. Spectroscopic determinations of the metallicity distribution have been clearer. The metallicity distribution is found to peak between [Fe/H] = --0.4 and --0.7 dex \citep{MBB+05,SBM+06,CMC+07,CBG+10,CMCD+12}. However, a high-metallicity tail extends toward solar metallicity \citep{SBM+06} and a low metallicity tail down to the metallicity of M54 \citep{CBG+10}.

{\bf Metallicity gradient:} spectroscopic and photometric studies have also shown a significant metallicity gradient. While the remaining Sgr dSph core is fairly well-mixed \citep{BFB99a}, there is an absence of the metal-rich populations in the tidal tails, which typically have metallicities around [Fe/H] = --1.1 dex \citep{CMC+07,CMCD+12}. This suggests selective stripping of metal-poor stars for the last $\sim$3 Gyr \citep{BNC+06,CMC+07}.

{\bf $\alpha$-element distribution:} spectroscopic surveys have also shown a smooth trend of decreasing [$\alpha$/Fe] with increasing [Fe/H], with the variation at each metallicity being dependent mostly on the study concerned than within each dataset. Reported values are between $\approx$+0.3 to +0.2 dex in the metal-poor stars, $\approx$+0.1 dex to --0.1 dex in the metal-intermediate population, and $\approx$--0.2 dex to --0.3 dex by the time the stars reach solar metallicity \citep{MBB+05,SBM+06,CBG+10}.

{\bf Age distribution:} \citet{LS97} and SDM+07 compiled age estimates of the populations as well as presenting their own. The metal-poor population is thought have an age of 9--12 Gyr, with estimates favouring the later value. The metal-rich population is generally found to be 4--6 Gyr old, though \citet{BFB99b} find the star-formation rate to have peaked 8--10 Gyr ago. The age and nature of the younger populations is more speculative, with star formation generally thought to have continued at a lower rate until less than 1 Gyr ago.

Photometric data containing only post-main-sequence stars suffers from considerable degeneracy between [Fe/H], [$\alpha$/Fe] and age. With the photometric data considered here, we cannot solve for all three simultaneously. However, if we are prepared to adopt the [$\alpha$/Fe] versus [Fe/H] and [Fe/H] versus age relationships given by the above literature, we can estimate the metallicity distribution of stars in our field by comparison to stellar isochrones.

Further degeneracy exists between age and distance, and between age, metallicity and reddening. Reddening over the field is well-constrained to be $E(B-V) = 0.15 \pm 0.05$ mag, which gives $E(J-K_s) \approx 0.075 \pm 0.025$ mag and $A_Ks = 0.06 \pm 0.02$ mag, where these values represent the mean and typical range across the field estimated from COBE dust maps (see Appendix for more details). The Sgr dSph is behind any likely reddening sources, which are expected to arise in the local Galactic Plane. We adopt 25 kpc as the distance of the Sgr dSph.

Finally, uncertainties exist in the isochrones themselves. Different isochrones, using different stellar models and evolutionary codes produce marginally different results. In older populations, one significant uncertainty is post-main-sequence mass loss. Even canonically-identical stars, such as those in globular clusters, have been found to experience significantly different post-main-sequence mass-loss histories \citep{CDOY+13}. Isochrones also treat mass loss differently: for example, the Padova models \citep{MGB+08} assume a formulism based on \citet{Reimers75}, but the actual RGB mass loss is only experienced by the models instantaneously at the RGB tip, though they allow for reddening of the RGB tip due to circumstellar dust\footnote{Whether significant numbers of RGB stars produce dust is not well determined over a wide range of stellar parameters. In $\omega$ Cen which, as a supposed tidally-stripped dwarf galaxy nucleus, may best represent the core of the Sgr dSph, we have found some evidence of RGB dust production \citep{MvLD+09}. However, this is questionable as it suffers from the same problems arising in a recent exchange of papers with Origlia et al.\ \citep{ORF+07,BvLM+10,ORF+10,MBvL+11,MBvLZ11,MSS+12}. These papers concern the globular cluster 47 Tuc, often used as a comparison to the Sgr dSph as it has a similar metallicity. In the McDonald et al.\ papers, confirmed by Momany et al., while we find dust production extends a bolometric magnitude below the RGB tip, dusty stars in this region are consistent with being on the AGB.}. We use the Dartmouth synthetic isochrones and horizontal branch tracks \citep{DCJ+08} for preference as they generally provide a good fit to stars in our age--metallicity regime (e.g.\ \citealt{MvLD+09}).

\begin{figure}
\centerline{\includegraphics[height=0.47\textwidth,angle=-90]{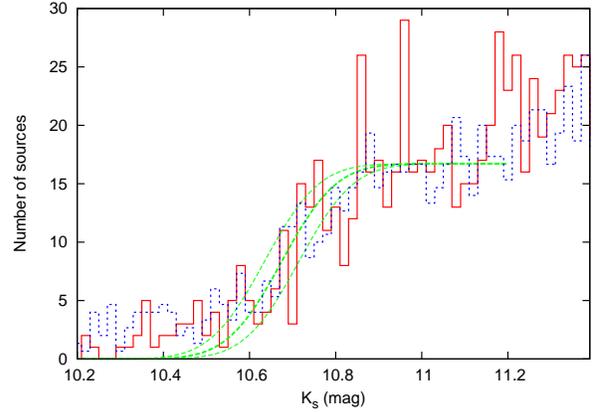}}
\caption{Histogram of VISTA $K_s$-band magnitudes of sources with colours matching those of Sgr dSph stars. The blue, dashed line shows the equivalent histogram for 2MASS $K_s$-band magnitudes (which cover a larger area) divided by 1.5. Cumulative Gaussian distributions are shown as dashed green line, representing the distribution expected for a step-function RGB tip smoothed by photometric variations of $\sigma$ = 0.1 mag. The thicker curve shows the adopted $K_s$-band 10.68 mag, while the thinner curves are placed at $\pm$0.04 mag.}
\label{SgrRGBHistFig}
\end{figure}

Figure \ref{SgrIsoFig} shows the extracted colour--magnitude diagram for the Sgr dSph, overplotted with Dartmouth synthetic isochrones and horizontal branch tracks \citep{DCJ+08}. We have included here isochrones with ages, [Fe/H] and [$\alpha$/Fe] to broadly represent the population catagorisations of SDM+07, but incorprating [$\alpha$/Fe] ratios from the recent literature cited above:
\begin{itemize}
\item In red, to represent the M54 population, an isochrone at                                 13 Gyr,        [Fe/H] = --1.7 dex,           [$\alpha$/Fe] = +0.2 dex;
\item In green, to represent the outer, metal-poor population, isochrones at                   10 and 12 Gyr, [Fe/H] = --1.3 dex,           [$\alpha$/Fe] = +0.2 dex;
\item In blue, to represent the metal-intermediate population, isochrones at                   4 and 8 Gyr,   [Fe/H] = --0.7 dex,           [$\alpha$/Fe] = 0 dex;
\item In magenta, to represent the metal-rich population, isochrones at                        2 and 3 Gyr,   [Fe/H] = 0 dex,               [$\alpha$/Fe] = --0.2 dex (note that the $\alpha$-depletion and age differences render these isochrones almost as blue as the metal-intermediate population);
\item In cyan, to represent the suspected very-metal-rich population, an isochrone at          1 Gyr,         [Fe/H] = +0.5 dex,            [$\alpha$/Fe] = --0.2 dex (note that this isochrone does not reach observed RGB tip).
\end{itemize}
Comparative theoretical HB tracks are shown at estimates zero-age HB masses, smoothed at either end by Gaussian of width 0.05 M$_\odot$. In the same order and colours, these are:
\begin{itemize}
\item $\!$\ [Fe/H] = --1.7 dex,          [$\alpha$/Fe] =  +0.2 dex, $m_{\rm ZAHB}$ = 0.60 M$_\odot$;
\item $\!$\ [Fe/H] = --1.3 dex,          [$\alpha$/Fe] =  +0.2 dex, $m_{\rm ZAHB}$ = 0.65 and 0.70 M$_\odot$;
\item $\!$\ [Fe/H] = --0.7 dex,          [$\alpha$/Fe] =     0 dex, $m_{\rm ZAHB}$ = 1.10 M$_\odot$;
\item $\!$\ [Fe/H] = 0 dex,              [$\alpha$/Fe] = --0.2 dex, $m_{\rm ZAHB}$ = 1.15 M$_\odot$;
\item $\!$\ [Fe/H] =  +0.5 dex,          [$\alpha$/Fe] =     0 dex, $m_{\rm ZAHB}$ = 1.95 M$_\odot$. Note that [$\alpha$/Fe] = --0.2 dex models are not available for this metallicity.
\end{itemize}

Figure \ref{SgrIsoFig} broadly confirms that isochrones representing the populations described by SDM+07 model the Sgr dSph galaxy well. The most-metal-poor population, corresponding to M54, is sufficiently low density and sufficiently blue to be poorly-distinguishable from the reddest, most-highly-extincted Bulge population. However, redward of this, there is a continuous distribution of stars. This distribution has a distinct peak towards redder colours where the higher metallicities, forming the bulk population, are expected. The sharpness of the red edge of the giant branch is amplified by the known [$\alpha$/Fe] versus [Fe/H] anti-correlation.

\subsubsection{Distance to the Sgr dSph}
\label{IsoSgrDSect}

The distance to the Sgr dSph cannot be constrained by our observations to better than the literature values, as we are limited in the number of clear evolutionary stages at which we can fix the various properties of the isochrones. Our best constraint is provided by the $K_s$-band magnitude of the RGB tip, which we fit to lie at $K_s = 10.68 \pm \sim$0.04 mag (Figure \ref{SgrRGBHistFig}). Our applied Gaussian spread in the RGB tip positions ($\pm$0.10 mag) allows for a depth to the galaxy of $\sigma_d < 4.7$ per cent of its distance, which will we expect to reduce substantially once intrinsic variation, differential reddening and photometric variability are taken into account. This equates to $\ll$1.2 kpc at the distance of the Sgr dSph.

The RGB tips of the Dartmouth isochrones representing the metal-intermediate population, which the above spectroscopic studies find to represent the bulk of the population, lie at $K_s = 10.80$ and 10.70 mag (for 4 and 8 Gyr, respectively) if we assume a distance of 25 kpc. Those representing the metal-rich population have a wider tolerance, lying at $K_s = 11.05$ and $10.51$ mag (for 2 and 3 Gyr, respectively). A comparison using the independent {\sc parsec} isochrones \citep{BMG+12} shows their corresponding isochrones at $K_s = 10.81$, 10.64, 10.96 and 10.43 mag, respectively (note that the {\sc parsec} isochrones do not allow non-solar [$\alpha$/Fe] ratios), suggesting that the choice of isochrones can add at most a $\pm$0.10 mag uncertainty to the distance. The uncertainty in the mean differential reddening (taken to be $A_Ks = 0.13$ mag) is very small and can be neglected compared to these errors.

Combining these magnitudes, giving double weight to the more-common metal-intermediate ([Fe/H] = --0.7) population, we form an expected range of 10.74 $\pm$ 0.19 mag for the position of the isochrone RGB tip at 25 kpc. Scaling these to the observed value, we find a distance to the Sgr dSph of 24.3 $\pm$ 2.3 kpc to the Sgr dSph. This is consistent with most previous estimates, but does not agree with the \citet{SML+11} result of 28--30 kpc.


\subsubsection{Spatial distribution of the respective populations}
\label{IsoSgr2Sect}

\begin{figure*}
\centerline{\includegraphics[height=0.47\textwidth,angle=-90]{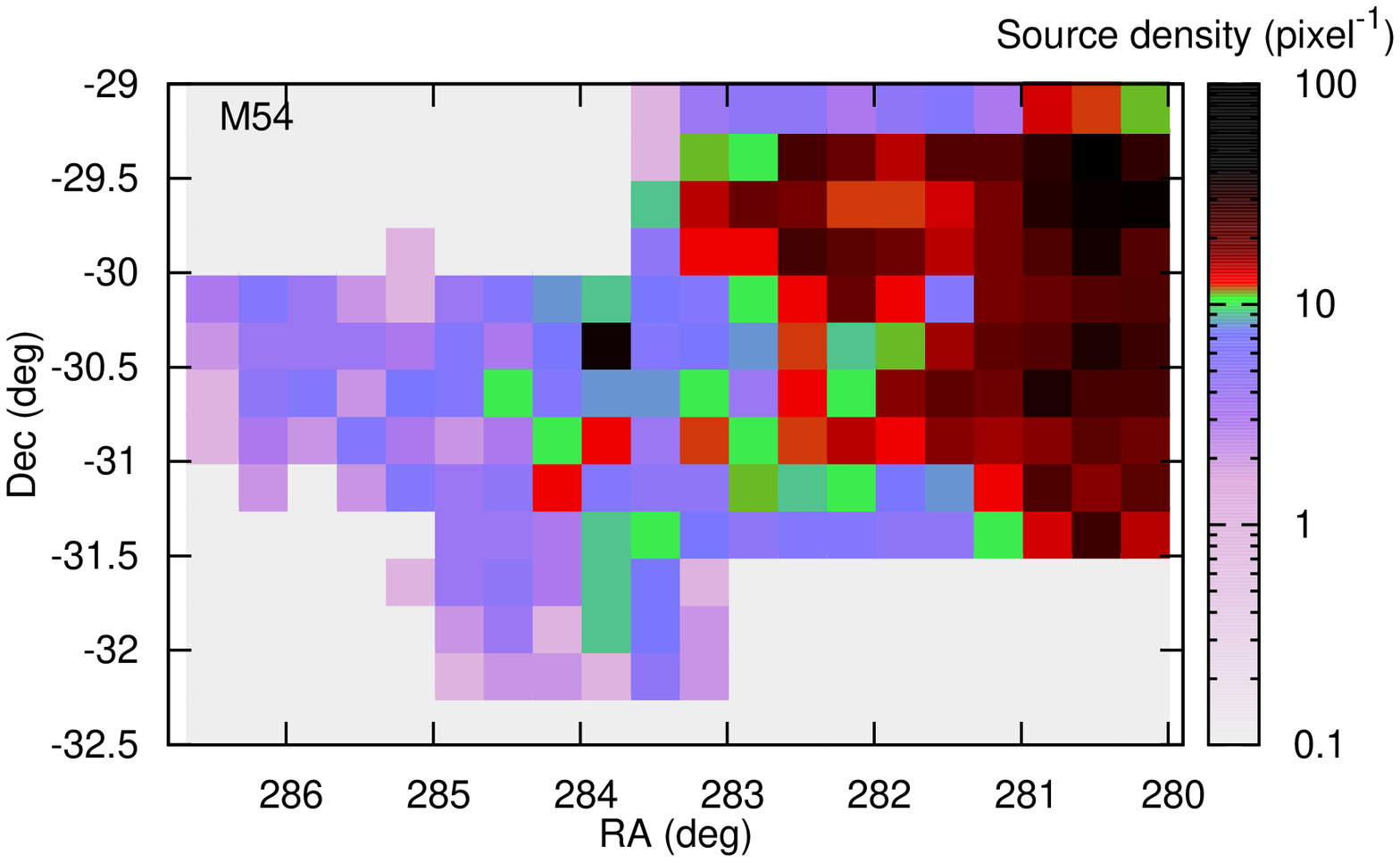}
            \includegraphics[height=0.47\textwidth,angle=-90]{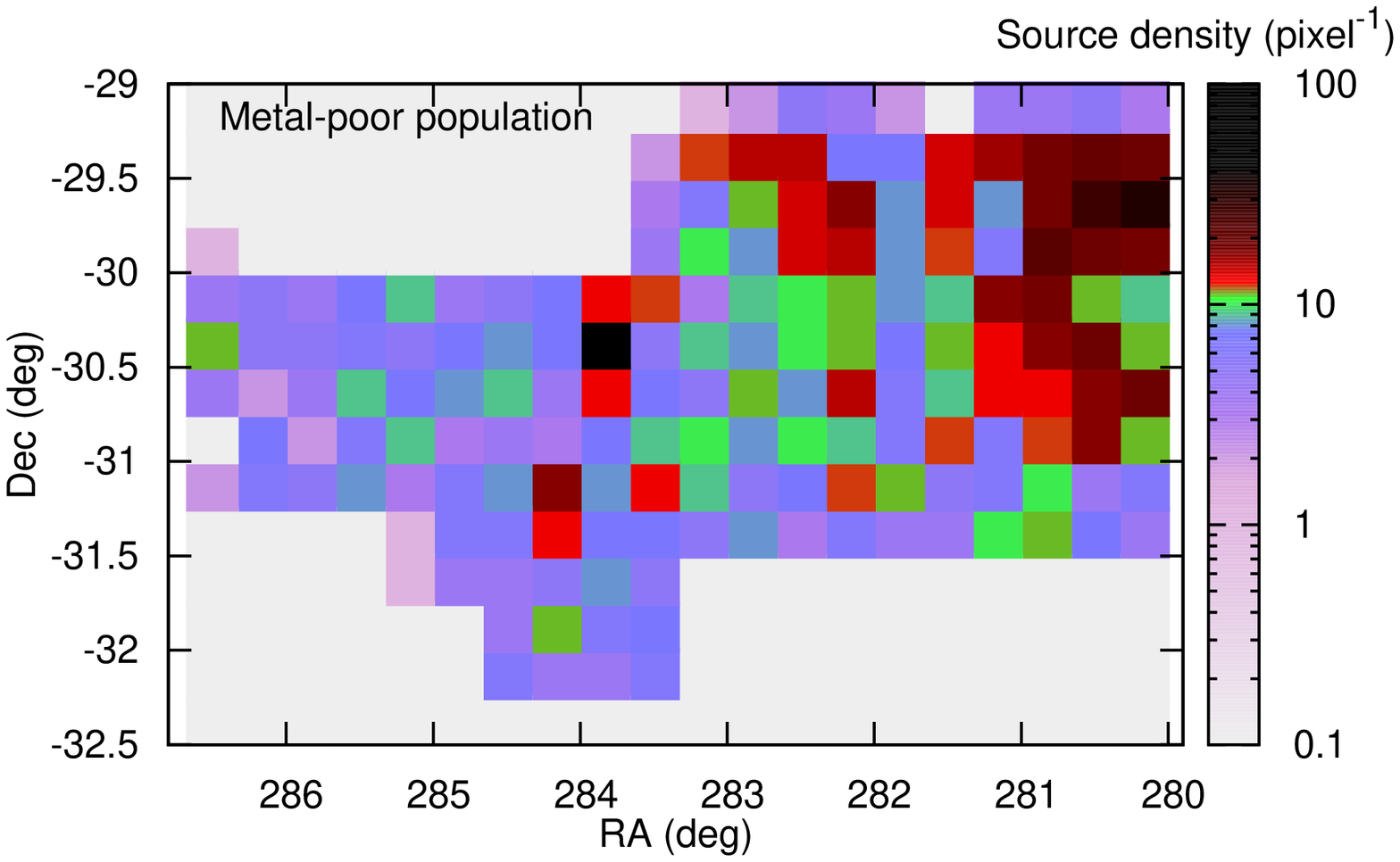}}
\vspace{-2mm}
\centerline{\includegraphics[height=0.47\textwidth,angle=-90]{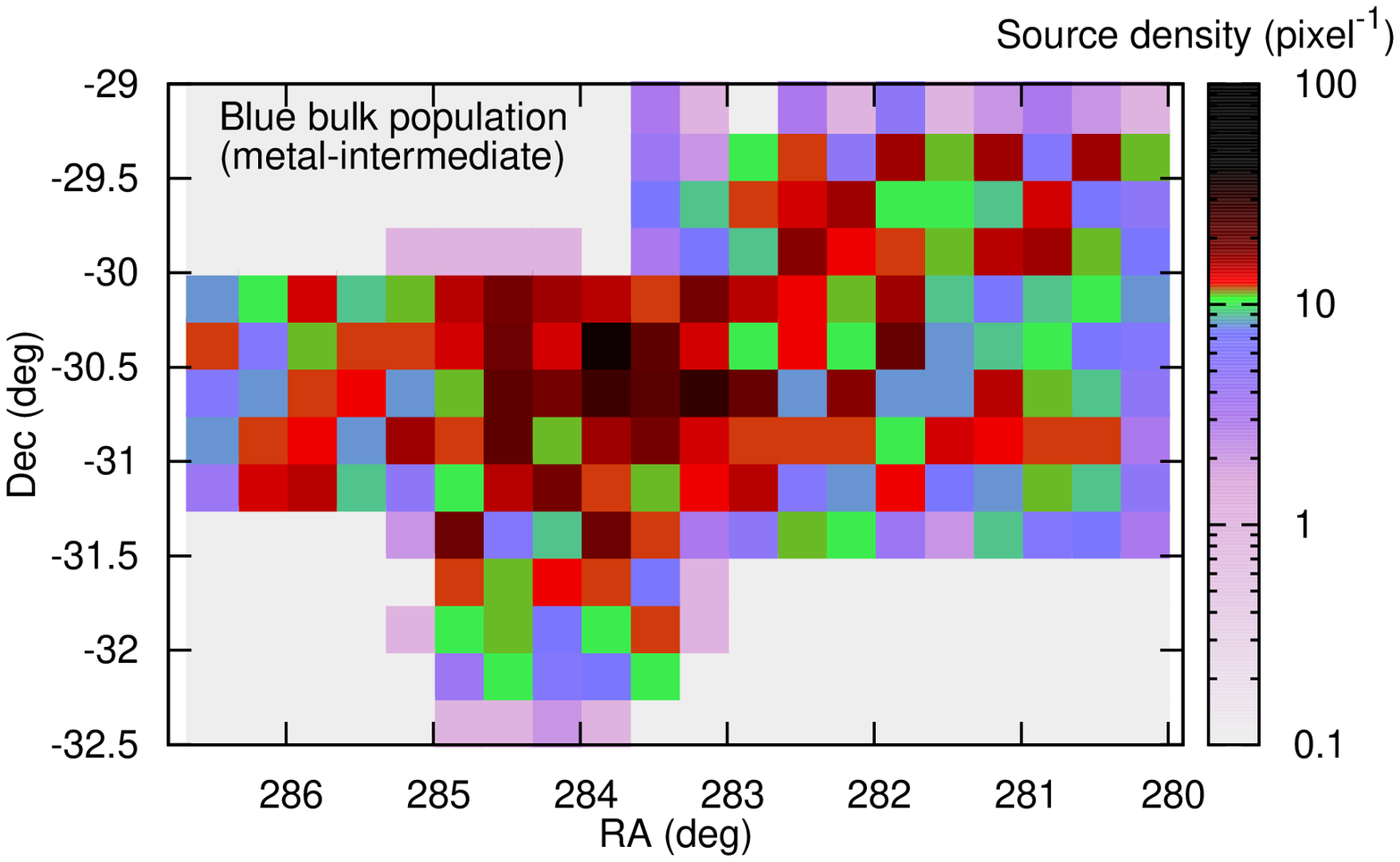}
            \includegraphics[height=0.47\textwidth,angle=-90]{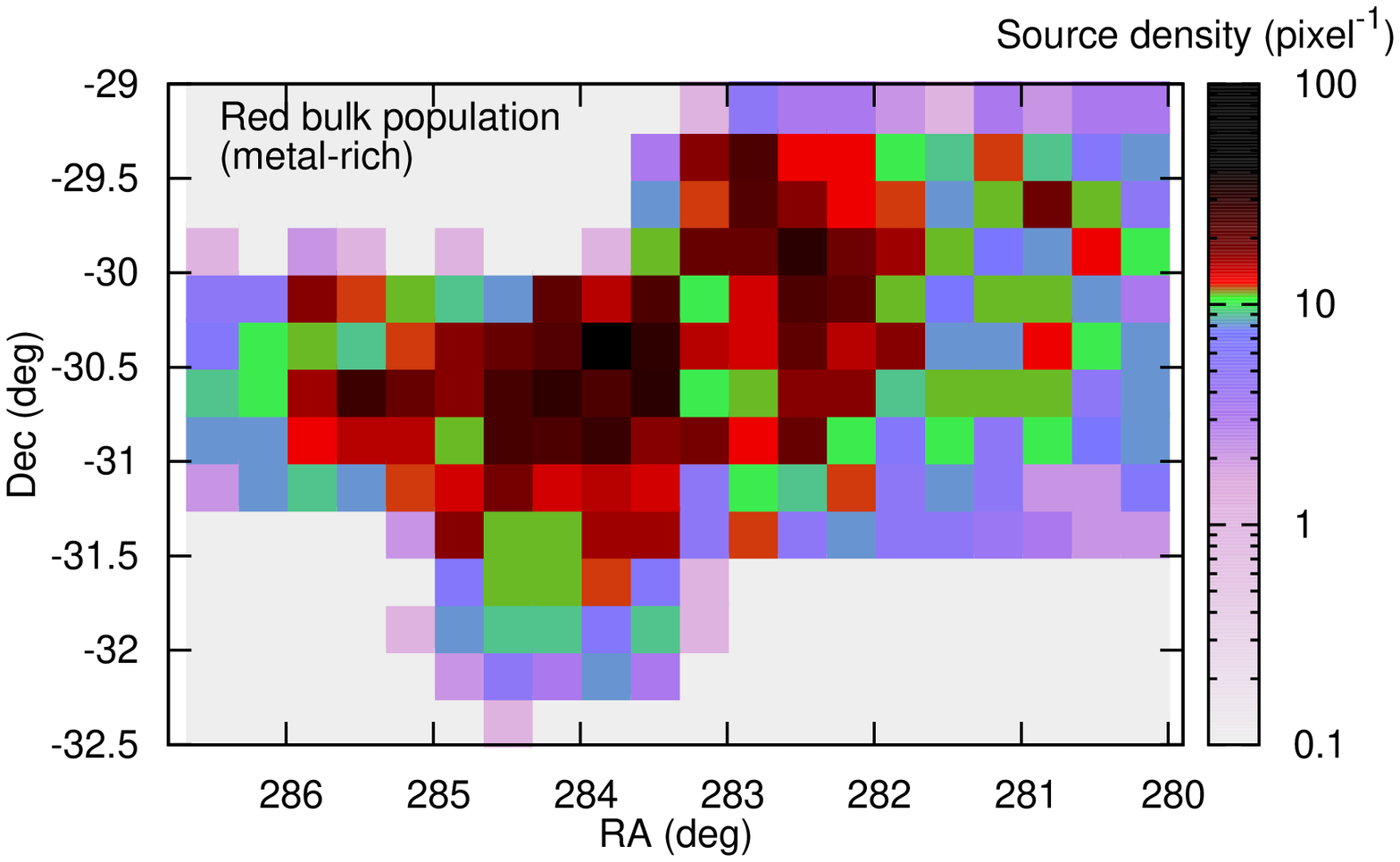}}
\vspace{-2mm}
\centerline{\includegraphics[height=0.47\textwidth,angle=-90]{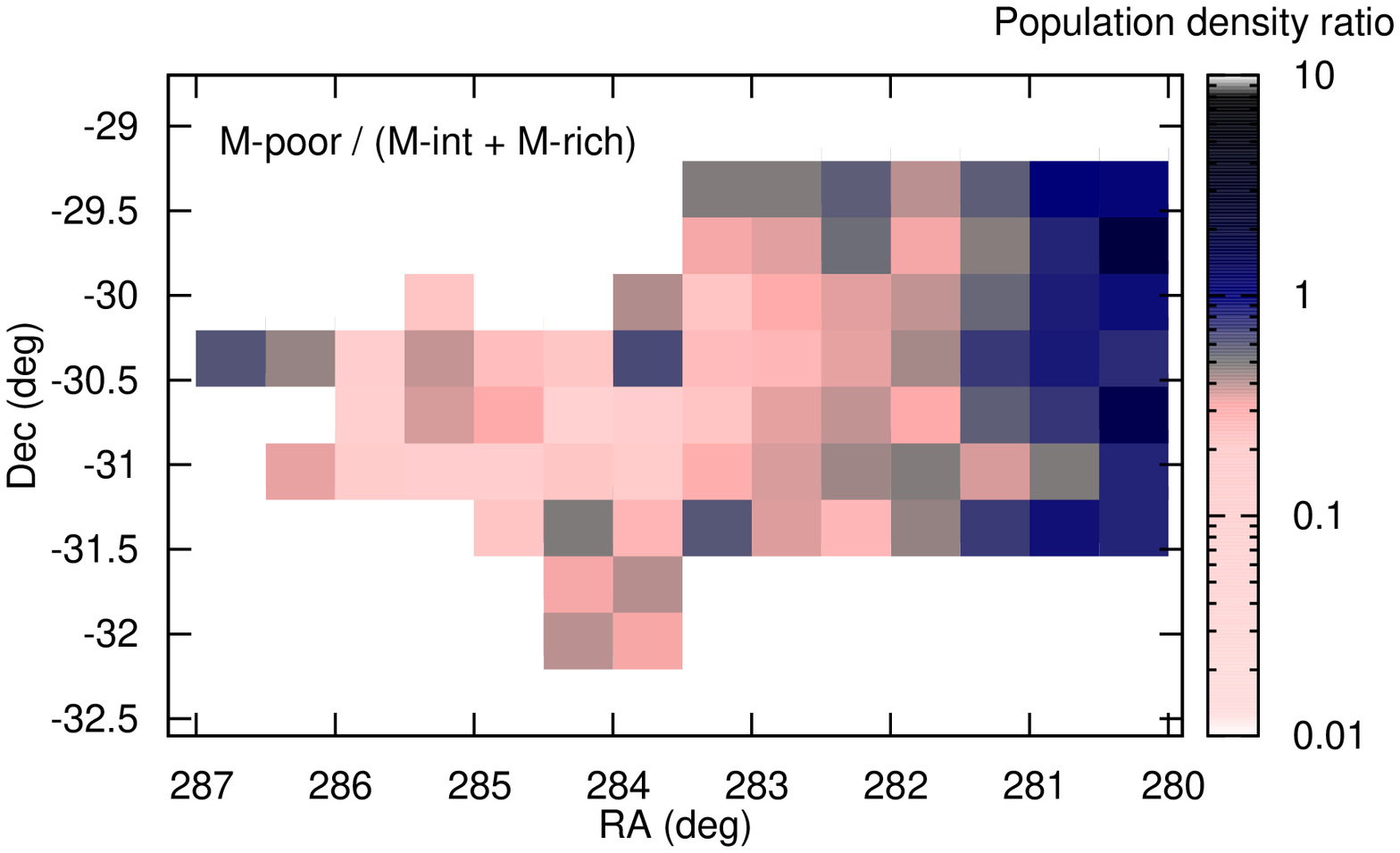}
            \includegraphics[height=0.47\textwidth,angle=-90]{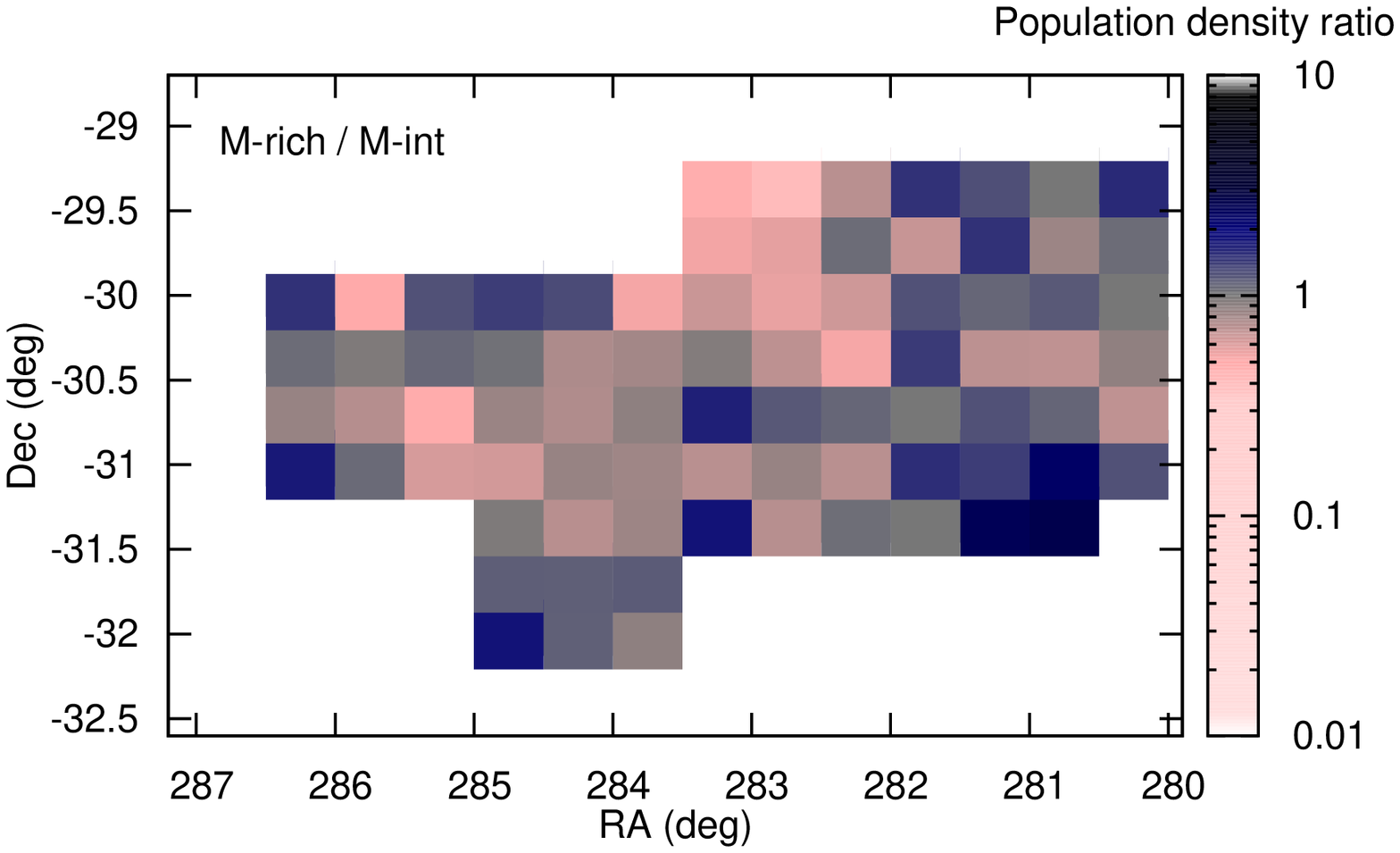}}
\caption{Top and central panels: the distribution of the four populations highlighted in Figure \ref{SgrIsoFig}. Note the incursion of Bulge stars into the M54 population and, to a lesser extent, the metal-poor population. Bottom panels: the ratio of the densities of (left, affected by the Bulge at low RA) the metal-poor population to the bulk metallicity population and (right) the metal-intermediate and metal-rich components of the bulk population. Ratios are only calculated where both numerator and denominator contain more than five stars. A colour version of this figure is available in the online edition of this work.}
\label{MetalMapFig}
\end{figure*}

\begin{figure*}
\centerline{\includegraphics[height=0.97\textwidth,angle=-90]{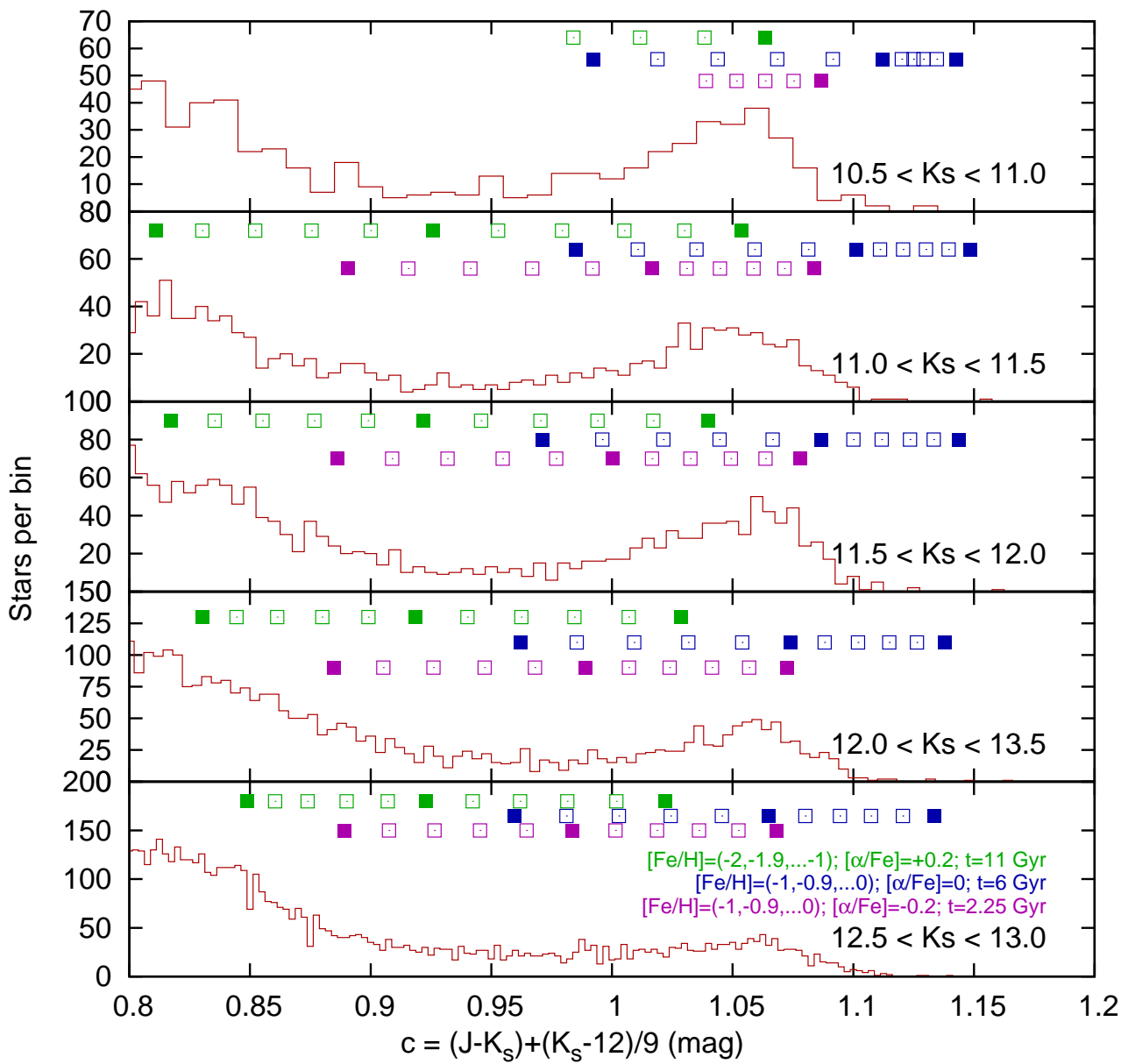}}
\caption{Colour distribution of stars on the Sgr dSph giant branch. For each 0.5-magnitude bin, a histogram of object colours is shown, where the formulism on the abscissa has been used to straighten the giant branch to avoid broadening it. The contribution around $c = 0.8$--0.9 mag is the Bulge giant branch; the contribution around $c = 1.0$--1.1 mag is the Sgr dSph. Points representing the colours of Dartmouth isochrones at that magnitude are also shown: green shows the metal-poor population; blue shows the metal-intermediate bulk population and magenta shows the metal-rich tail of the bulk population. Isochrones are presented in steps of $\Delta$[Fe/H] = 0.1 dex with filled points every 0.5 dex. Only the metal-rich isochrones are shown in the top-most plot, as the metal-poor isochrones do not reach these magnitudes.}
\label{ColourHistFig}
\end{figure*}

Figure \ref{MetalMapFig} shows the distribution of the different populations highlighted in Figure \ref{SgrIsoFig}. Note that we have not generated a region for the expected location of the most-metal-rich population as the isochrones at this metallicity are poorly calibrated. The population belonging to M54 is heavily contaminated by stars from the Bulge, which cluster in the direction of the Galactic Centre (the top-right of the maps). It is not possible to meaningfully extract the distribution of stars originated in M54, which appears as a high-density (black) dot without extensive radial velocity surveys. However, the density of M54 sources to the left of the map is much less than any of the subsequent maps, showing a general absence of any population of M54 stars outside of the cluster itself. The population of M54 stars mixed in with the rest of the Sgr dSph galaxy must be a very small fraction of the total number of stars.

While \citep{SML+11} place M54 in front of the Sgr dSph galaxy, this disagrees with the consensus opinion that they are at the same distance and M54's projected spatial co-incidence with the Sgr dSph core implies that it is the galaxy's nucleus \citep{MBFP05,BIC+08}. This is despite its unusually-low metallicity for a galactic nucleus, and it is presumed to have tidally decayed into its current position \citep{BIC+08}. The lack of an obvious M54 population in the rest of the Sgr dSph galaxy implies there has been little dispersion of the M54 population or mixing of it into the rest of the galaxy. One might expect such mixing if M54 arrived at the Sgr dSph core via tidal accretion or a multiple-body encounter, as we see with Sgr dSph stars being tidally dispersed into the Milky Way's halo. While M54's low metallicity would appear to prevent it forming \emph{in situ}, we would expect that the passage that brought it to its current position is a dynamically-simple, low-velocity merger that does not cause significant tidal stripping. We hypothesise that the current Sgr dSph nucleus may have simply built up around the gravitational potential of M54.

The metal-poor population, approximately representing metallicities from [Fe/H] = --1.6 to --0.9 dex, is also slightly contaminated by Bulge stars. However, regions at higher Galactic latitude (left and bottom of the maps) show more of these metal-poor stars than the M54 population, indicating that the metal-poor population is better spread throughout at least the high-Galactic-latitude end of the Sgr dSph.

The metal-intermediate and metal-rich populations both show a roughly ellipsoidal shape, centred around M54, with very little obvious contamination from the Bulge. As with the other populations, the bulk population shows a central point centered on M54 itself. This central cusp was shown by \citet{MBFP05} to be a few arcminutes in diameter. It overlies a broader distribution of stars, which falls to half the central density at approximately two degrees from M54 in Galactic latitude and one degree from M54 in Galactic longitude.

The relative ratios of the metal-rich to metal-intermediate components (i.e.\ the blue and red sides of the visually-obvious Sgr dSph giant branch) show little variation with position (Figure \ref{MetalMapFig}, bottom-right panel). The differences in star count that do exist mostly trace the small differences in interstellar reddening in the \emph{IRAS} dust maps \citep{SFD98}. The ratio between metal-poor and bulk metallicity (metal-intermediate and -rich) populations shows more structure, although some of this is attributable to contamination from the Galactic Bulge in the metal-poor sample. Even when only considering the left-hand side of the map, away from the Bulge, there is still a general trend whereby pinker colours, representing a greater fraction of metal-rich stars, concentrate around M54 (though not in the pixel containing M54 itself). This indicates a metallicity gradient in the galaxy. Further study at larger radii would be helpful when the VISTA Hemispheric Survey completes its catalogue of this area.

\subsubsection{Metallicity distribution of the surveyed region}
\label{IsoSgr3Sect}

The four colour--magnitude regions, in order of increasing metallicity, contain 2005, 1565, 1893 and 2031 stars. The most-metal-poor region, representing M54, is dominated by Bulge stars (Figure \ref{MetalMapFig}). We can crudely estimate from their spatial bias ($B$) towards the Galactic Bulge that $>$64 per cent of stars in this colour--magnitude region belong to the Bulge, thus we discount it from our analysis.

Most of the 1565 objects in the metal-poor region are Sgr dSph stars, though this population also suffers from some contamination. When compared to the two metal-rich populations, we find the metal-poor population contains at most 29 per cent of the Sgr dSph's giant stars (note that, for the mass range we are exploring, the rate of evolution is roughly mass-invariant on the giant branch; e.g.\ \citealt{DCJ+08}). In reality, the number is likely to be somewhat less, due to the contribution from the Bulge, and potential contamination in the metal-poor region by common-envelope binaries and AGB and post-AGB stars from the bulk population.

Figure \ref{ColourHistFig} shows the colour distribution of stars on the Sgr dSph giant branch, along with the corresponding colours expected from Dartmouth isochrones representing the three remaining populations. We here skew the colour--magnitude diagram using a colour term, $c = (J-K_s)+(K_s-12)/9$, which allows the observed Sgr dSph giant branches to have a roughly constant value of $c \approx 1.06$ mag: the ($J-K_s$) colour of the giant branch at $K_s = 12$ mag. The isochrones have been chosen to match the three observed populations, at [$\alpha$/Fe] = +0.2, 0 and --0.2 dex with ages of 11, 6 and 2.25 Gyr, but are shown at a variety of different metallicities from [Fe/H] = -2 to -1 dex for the metal-poor stars, and [Fe/H] = -1 to 0 dex for the metal-intermediate/-rich components.

The degeneracy between [Fe/H] and both [$\alpha$/Fe] and age can be seen by comparing the colours given by the three sets of isochrones at [Fe/H] = -1. For identical metallicities, the isochrones show a range in colour of $\Delta c \approx \Delta(J-K_s) \approx 0.15$ mag. This degeneracy means we are unable to recover the metallicity distribution of the Sgr dSph without much more precise constraints on the absolute relationship between age and metallicity, and between [$\alpha$/Fe] and [Fe/H].

\begin{figure}
\centerline{\includegraphics[height=0.47\textwidth,angle=-90]{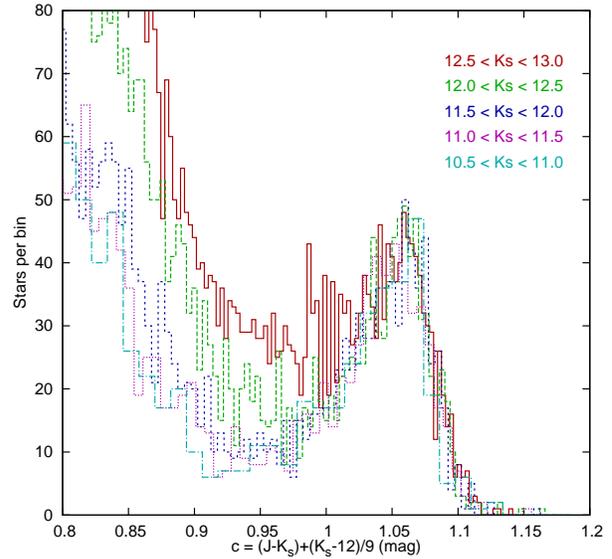}}
\caption{Colour distribution of stars on the Sgr dSph giant branch. As Figure \ref{ColourHistFig}, but with histograms overplotted and bin size matched to equate star counts on the Sgr dSph giant branch.}
\label{OverplotHistFig}
\end{figure}

That is not to say that we cannot make inferences from this diagram, but that they must remain as such and should not be taken as direct evidence in their own right. We note that the strength and coherence of the Sgr dSph giant branch histogram peak is lost at fainter magnitudes, and that most stars occupy the space between $0.9 < c < 1.0$ mag where, canonically, the metal-poor population is expected to lie. This is not an artifact of decreasing signal-to-noise: stars in all these regions have roughly similar photometric errors, and the red side of the giant branch falls off at the same rate at faint magnitudes as at bright magnitudes. This can mostly be explained by the increasing contribution of Bulge stars at bluer colours. Bulge stars make a significant contribution up to $c \approx 0.91$ mag in the brightest bin, where the histogram reaches a minimum. However, with decreasing brightness, this point moves progressively redwards in this colour space as Bulge stars increase in number, with their reddest significiant contribution in the subsequent bins being (from brightest to faintest) $c \approx 0.92$, 0.93, 0.96 and 1.00 mag. At magnitudes fainter than $K_s = 13$ mag, the Sgr dSph and Bulge begin to merge in the colour--magnitude diagram. Underneath this, the distribution of Sgr dSph stars remains largely unchanged (Figure \ref{OverplotHistFig}).

We can also infer that the Sgr dSph population should be slightly $\alpha$-element depleted. We have artificially removed the slope of the giant branch, but the isochrones have independent slopes. The younger isochrones at [$\alpha$/Fe] = --0.2 dex have a much more constant colour ($c$) than the middle-aged isochrones at [$\alpha$/Fe] = 0 dex. This slope is more defined by [$\alpha$/Fe] than age, hence the Sgr dSph population should have at least slightly sub-solar [$\alpha$/Fe]. However, this result depends strongly on the accuracy of the isochrones.

Finally, the narrowness of the Sgr dSph giant branch is surprising, when given the range of metallicities where significant numbers of stars are found spectroscopically ($--1 \lesssim [Fe/H] \lesssim 0$ dex; \citealt{CBG+10}, their figure 8). Based again on the accuracy on isochrones and ages involved, this implies an increase from [$\alpha$/Fe] = 0 dex at [Fe/H] $\approx$ --0.7 dex to [$\alpha$/Fe] = +0.2 dex by [Fe/H] $\approx$ --0.2 dex. This is a little greater than the spectroscopically-derived values visible in figure 20 of \cite{CBG+10}, but probably within a factor of two of the slope of their abundance-averaged values.

In summary, our observations are consistent with the current understanding of the Sgr dSph population, though a spectroscopic survey to determine the metallicities and $\alpha$-element enhancements of its stars would improve matters further. We anticipate the results of the Apache Point Observatory Galactic Evolution Experiment (APOGEE; \citealt{ZJF+13}) will meet these criteria.

\subsection{The Galactic Bulge}
\label{IsoBulgeSect}

\begin{figure*}
\centerline{\includegraphics[height=0.47\textwidth,angle=-90]{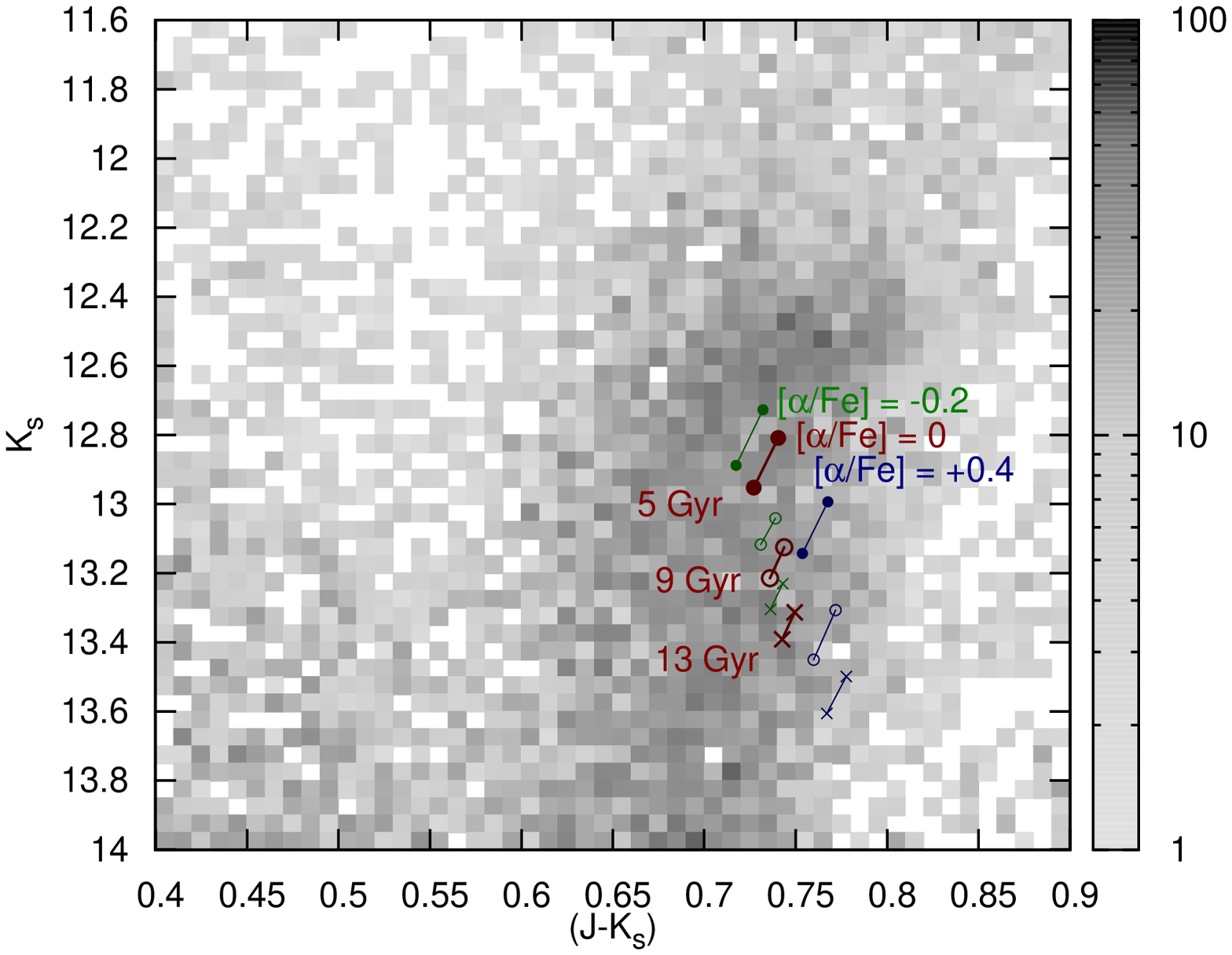}
            \includegraphics[height=0.47\textwidth,angle=-90]{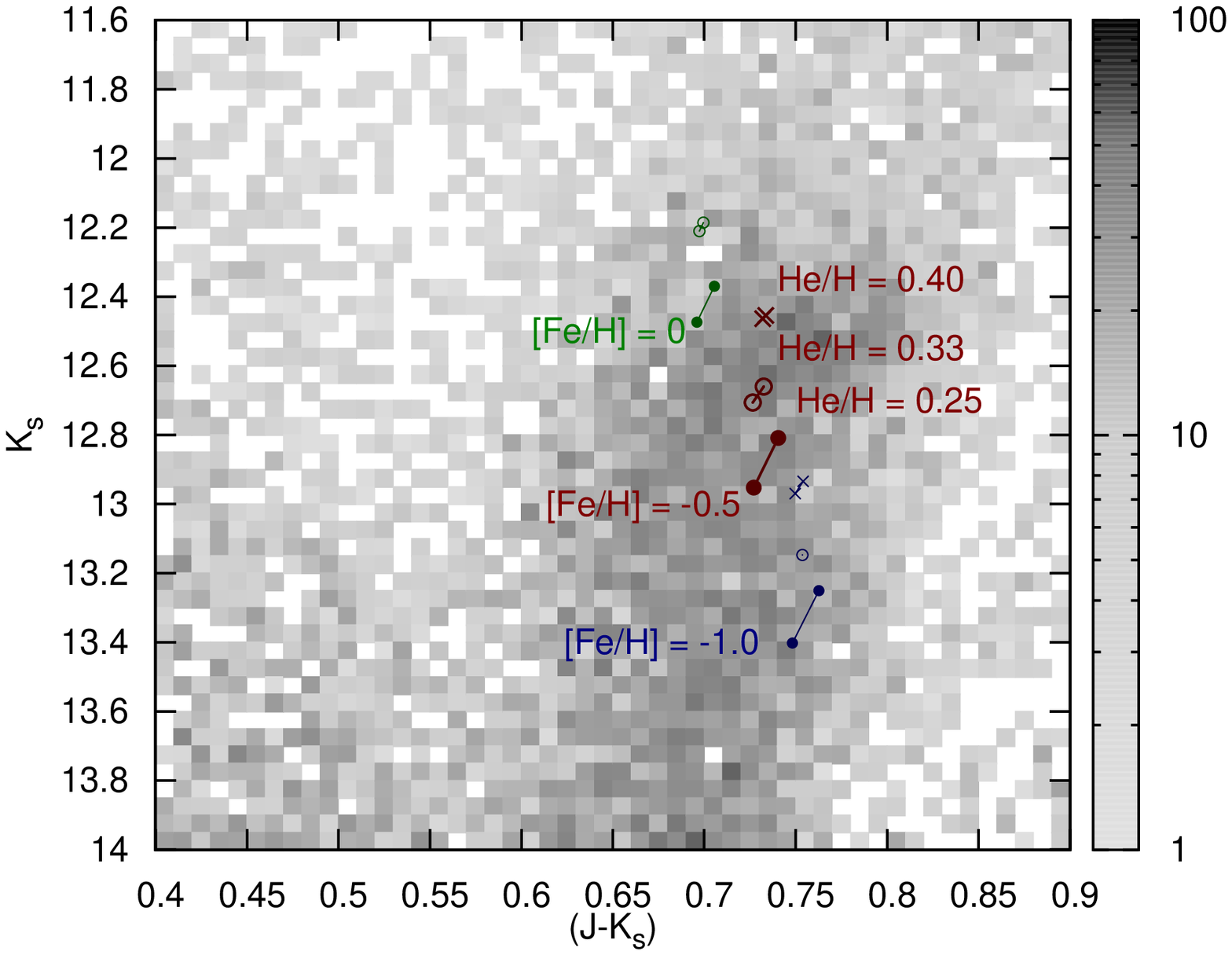}}
\vspace{-2mm}
\centerline{\includegraphics[height=0.47\textwidth,angle=-90]{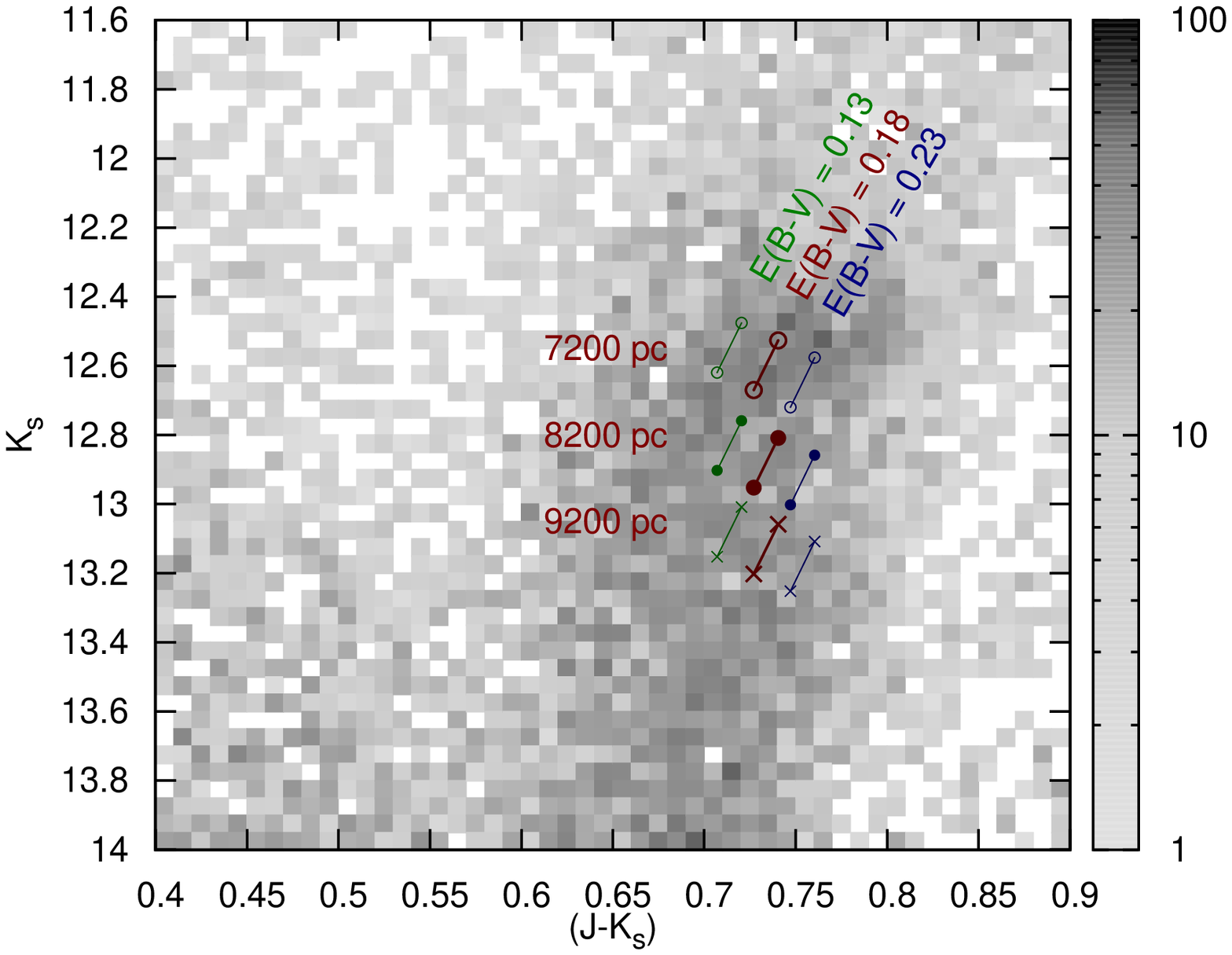}
            \includegraphics[height=0.38\textwidth,angle=-90]{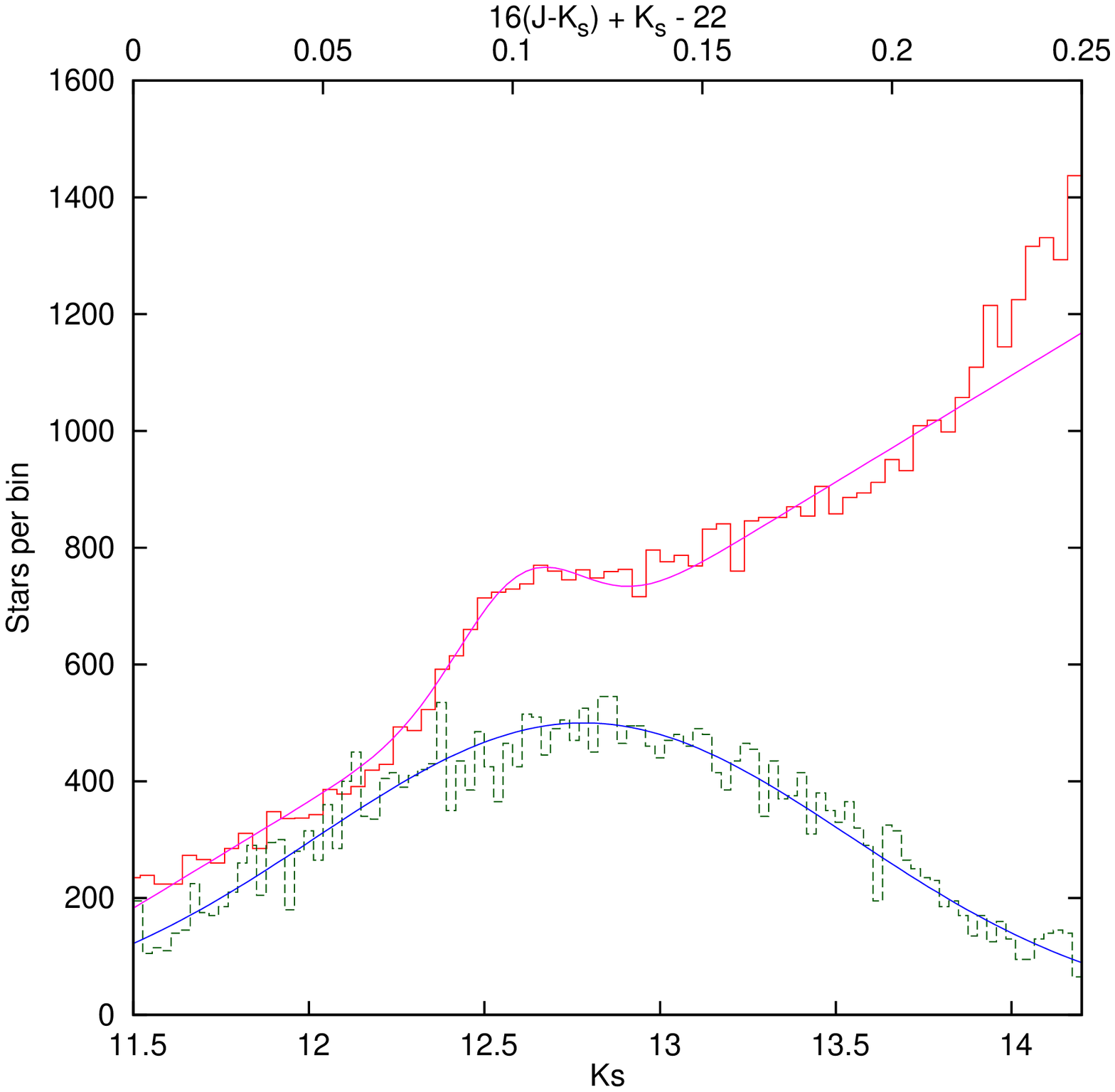}}
\vspace{-2mm}
\centerline{\includegraphics[height=0.47\textwidth,angle=-90]{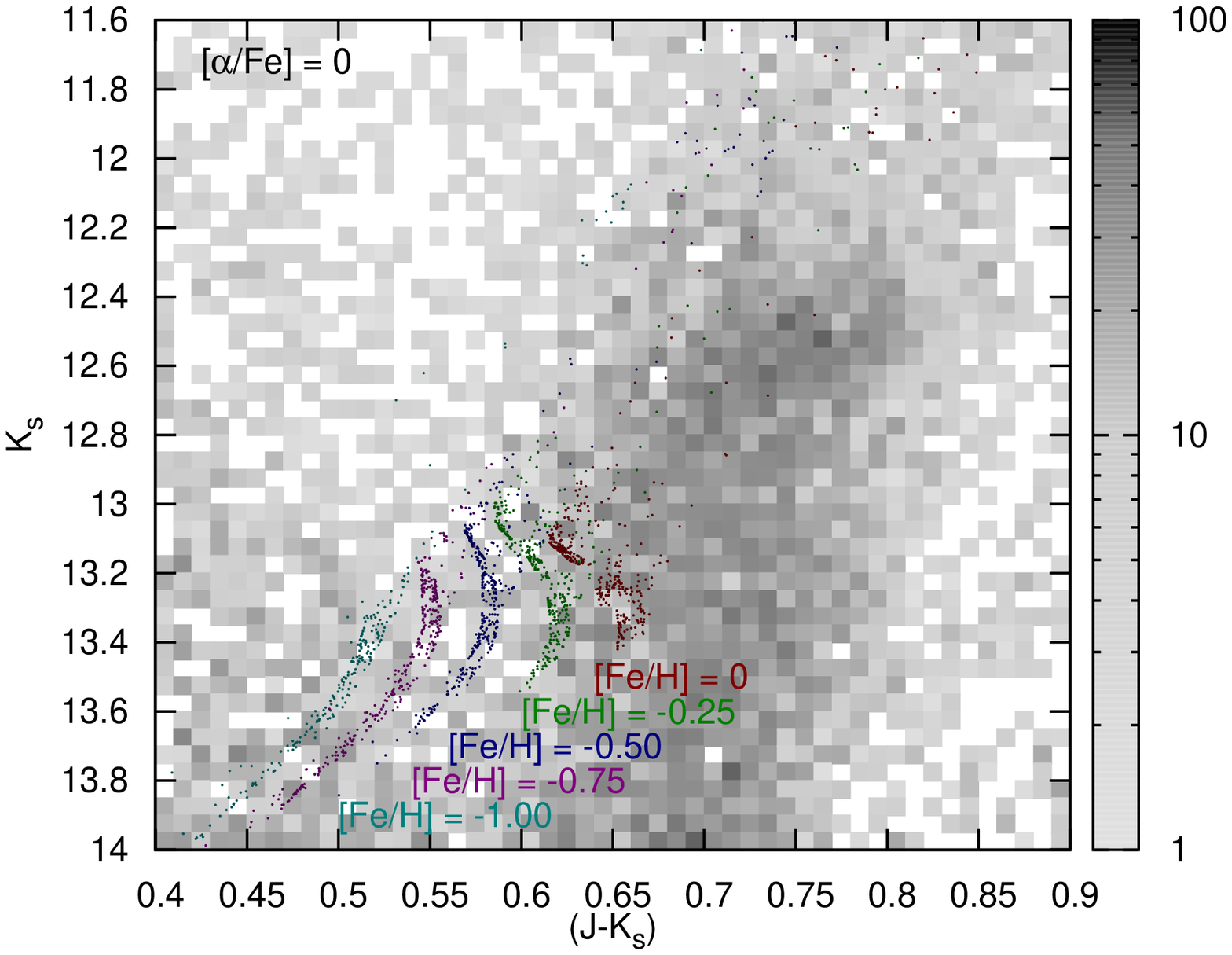}
            \includegraphics[height=0.47\textwidth,angle=-90]{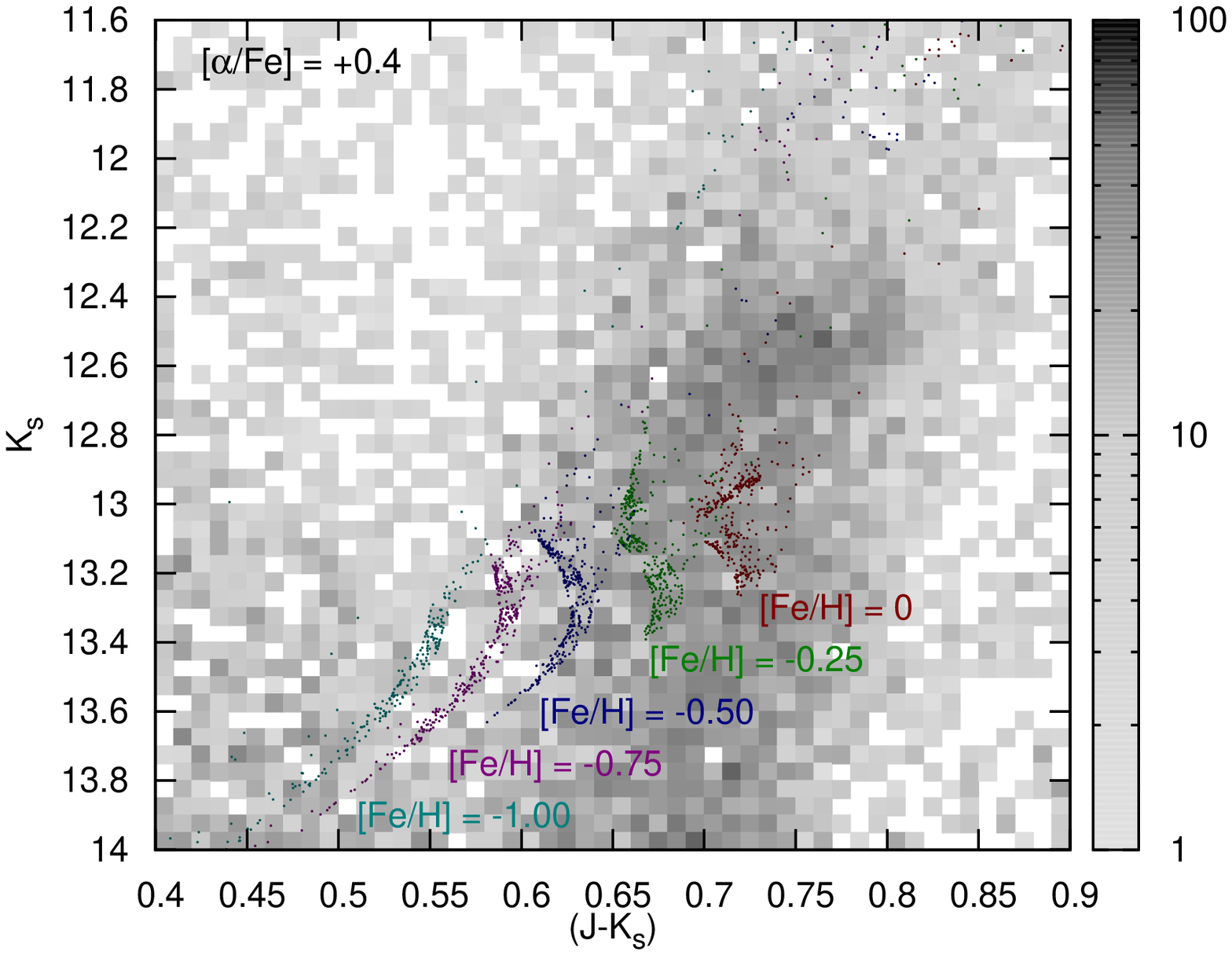}}
\caption{The statistically-extracted Bulge population from Figure \protect\ref{BiasFig}, with Dartmouth isochrones. The upper panels show the allowed variation of the RGB bump position with (top left) age, [$\alpha$/Fe]; (top right) helium fraction and global metallicity; (centre left) distance and reddening. Lines show the approximate extent of the bump (limited by the isochrone resolution), marked by points at either end. A distance of 8.2 kpc corresponds to our location being 8 kpc from the Galactic Centre. Centre-right panel: the upper line shows the luminosity function of stars in the region of the Bulge RGB bump (data: red, solid line; model: magenta, dotted line; following bottom axis). The corresponding colour distribution of those stars is also shown (data: green, long-dashed line; model: blue short-dashed line; following top axis). The colour distribution accounts for the slope of the giant branch in the colour-diagram, as defined in Figure \ref{CMDFig}, and is therefore not a pure $(J-K_s)$ colour. Bottom panels: horizontal branch models showing the effects of global metallicity and [$\alpha$/Fe] enhancement, where a range of stellar masses are plotted. A colour version of this figure is available in the online edition of this work. }
\label{BulgeFig}
\end{figure*}

The best-isolated group of Galactic Bulge stars is the RGB bump, which we show in Figure \ref{BulgeFig}. The RGB bump appears roughly Gaussian in our observations, and peaks near $K_s$ = 12.60 mags by $(J-K_s)$ = 0.71 and a Gaussian width of $\sigma_{Ks}$ = 0.25 mags and $\sigma_{(J-Ks)}$ = 0.10 mags.

We overplot in Figure \ref{BulgeFig} the locations of the RGB bump and horizontal branch in various Dartmouth isochrones. We begin here with an isochrone for solar-scaled abundances at [Fe/H] = --0.5, 5 Gyr old, placed at 8200 pc behind $E(B-V) = 0.18$ mags of reddening. The extinction of the average Bulge star is a little higher than that of M54. The IRAS+COBE/DIRBE dust maps \citep{SFD98} show that the Bulge-dominated regions of the survey area typically have $E(B-V)$ between 0.13 and 0.23 mag.

Figure \ref{BulgeFig} shows that a variety of degeneracies prevent us from getting accurate data about the typical Bulge star from this observation alone. However, we can constrain some parameters. Strong $\alpha$-element enhancements can be ruled out, as these would make the RGB bump too faint. Metallicities of [Fe/H] $\lesssim$ --0.5 dex are ruled out too: partly by the RGB bump luminosity, but also by the horizontal branch tracks, which appear to lie on top of the giant branch. Strong helium enhancements would also give rise to bluer horizontal branch models (not plotted here but see, e.g., \citealt{GCB+10}), so this is also disfavoured. Populations older than $\sim$9 Gyr are also disfavoured, as their RGB bumps are not bright enough. Closer distances are favoured, which we discuss below. Thus the dominant Bulge population along this line of sight is best fit by a population of roughly solar abundances which is relatively young (perhaps $\sim$5 Gyr or even less) and closer to us than the Galactic Centre. It should be noted that we have not taken into account errors within the isochrones here, which are expected to broadly match the $\pm$0.1 mag variations in $K_s$-band magnitude noted earlier (and are expected to be relatively less in ($J-K_s$)). We further note that we only examine the strongest signal from the Bulge. Populations with fewer stars or a wider spread of parameters will not leave such a strong signal in the colour--magnitude diagram.

Our observations do not provide strong evidence for the fainter red clump seen previously \citep{MZ10,SZM+11,NFA+12}. This has been attributed to a peanut- or X-shaped Bulge with a major axis aligned with our line of sight. Given our implication that the Bulge population we observe is closer than the Galactic Centre, we suggest its absence is due to missing the more-distant population due to projection effects. Our larger angle from the Galactic Centre and our positive offset in Galactic longitude ($l = 5.6^\circ$; cf.\ \citet{NFA+12}, their figure 5) may pass over the nearer component of the Bulge, but miss the far component. Alternatively, it may indicate an absence of this reputed component at greater distances from the Galactic Plane.

Our conclusions on the Galactic Bulge population would be much better constrained by [Fe/H] and [$\alpha$/Fe] measurements from spectra of Bulge stars in this line of sight. Differentiating the distances of the Bulge RGB bump stars would also place strong constraints on the population: distance measurements by \emph{Gaia} are expected to be close to the useful statistical limit for individual stars, but an ensemble of stars in the RGB bump could provide strong constraints on the distance of this population. Colour--magnitude diagrams showing the main-sequence turn-off would also be helpful in constraining the age, removing some of the degeneracy among the other parameters. We advocate a spectroscopic survey of Bulge stars in this field, combined with deep imaging to probe the main-sequence turn-off. The aforementioned APOGEE spectroscopic survey is anticipated to provide the necessary spectral coverage, while optical imaging has recently been acquired using the Blanco-4m Dark Energy Camera, which will be presented in a future publication.

\section{Conclusions}
\label{ConcSect}

Images of the densest 11 square degrees of the Sgr dSph at $Z$, $J$ and $K_{\rm s}$ were compiled into a catalogue of 2\,921\,920 sources, and cross-matched to other major optical and infrared catalogues in the literature. This catalogue was used to confirm the properties of the Sgr dSph and the foreground Galactic Bulge population.

Comparison to stellar isochrones confirms that the spheroidal Sgr dSph core is dominated by a population with [Fe/H] $>$ --1 dex, with a significant anti-correlation of [$\alpha$/Fe] and [Fe/H]. From the distribution of the galaxy's stars, we find it well represented by an oval shape measuring approximately 4$^\circ$ in Galactic latitude and 2$^\circ$ in Galactic longitude. We find some evidence for a metallicity gradient in the galaxy, and place it at 24.3 $\pm$ 2.3 kpc using isochrone fitting. Although we cannot accurately constrain the metallicity distribution of the Sgr dSph galaxy's stars, the metal-poor populations make up $\lesssim$29 per cent of the observed bright giants in the galaxy.

The RGB bump and helium-burning giant branch clump of the southern edge of the Galactic Bulge are matched with isochrones. A match is found only for isochrones representing a younger population, or one of particularly high metallicity, high [$\alpha$/Fe] and/or high He/H. We consider a younger population of near-solar composition most likely, possibly with some $\alpha$-element enhancement. We do not see evidence for a split RGB bump in the Bulge population, though this may be attributable to our line of sight.

The ongoing VVV Survey (\citealt{MLE+10,SHM+11}) is mapping the bulge regions with VISTA in the NIR, and will provide complementary information on the Sgr populations in the innermost regions.


\section*{Acknowledgements}

We are thankful to Mike Irwin, Mike Read and the staff at the Cambridge Astronomical Survey Unit and Vista Science Archive for their help and support in understanding the systematics associated with our data. Based on observations made with ESO telescopes at the La Silla Paranal Observatory under programme ID 089.D-0113 and data products from the Cambridge Astronomical Survey Unit. DM is supported by BASAL CATA Center for Astrophysics and Associated Technologies PFB-06, and by Project FONDECYT No. 1130196.


\appendix
\section{Merging \& calibration}
\label{CalibSect}

\subsection{Source matching across filters}
\label{MergeSect}

We have performed here the simplest algorithm to match sources across images taken in different filters, that of nearest-neighbour matching. The only parameter used here is that of the distance between two detections at which one is prepared to declare that they are the same object. This radius is determined here empirically. We have taken two contiguouD.~Minniti$^{3,4}$,s tiles observed at the same epoch in the same filter, and concatenated them into a single list of sources. The overlap region between the tiles contains some $\approx$8700 duplicated detections. We then attempt to match sources over the entire concatenated list using a range of different maximum radii.

We can identify the effectiveness of choosing a particular radius by comparing the number of false positive and false negative matches that we obtain. A proxy for false positives is relatively easy to determine: sources in non-overlapping regions of tiles should, by definition, only be observed once. If two objects are matched within a certain radius in these non-overlapping regions, we can define it as a false positive. This can then be scaled to the area of the overlap regions to estimate how many similar false positive matches we can expect in the overlapping areas. False negatives are a little more difficult to determine analytically, but can be approximated as the difference between the number of matches gained at that radius and the number gained at an arbitrarily large ($\sim$2$^{\prime\prime}$) radius after correcting for the number of false positives.

The number of matches is little affected by choice of radius, changing by $<$1\% between a separation of 0$\farcs$5 and 1$\farcs$4. We have adopted a conservative 0$\farcs$8 when matching sources. We estimate that $<$0.07\% of these matches will result in incorrect matching of detections between filters, though we may wrongly identify $\approx$0.4\% of detections as two different objects in this way. We have adopted a more-liberal 1$\farcs$2 when merging different tiles. We estimate that $\approx$0.11\% of sources merged in this way are actually two separate detections, while $\approx$0.16\% of objects should be merged but were not. While all percentages here are approximate, we note that these choices of radii were taken to make our catalogue more robust against poor photometry.

We have also examined the effectiveness of target identification by comparing our catalogue to that of 2MASS in a typical 2.25$^\circ$ $\times$ 1.30$^\circ$ region, away from the edges of the observed field, regions affected by a poorly-functioning VISTA chip, and M54, but including regions of overlapping tiles. Of the 102\,612 2MASS sources with photometric qualities AAA through CCC, we recover 102\,051 (99.45\%). Of the missing 561 sources, all but 62 have a quality flag of C in the 2MASS $K_s$ band, and only 43 have a 2MASS quality rating of AAA. Examination of these missing sources show that they are mostly detected, but outwith our 1$\farcs$2 matching radius. The difference in astrometry is likely to be due to blending by brighter objects in some cases, poor data quality in others, with a few nearby objects also affected by high proper motion.

\subsection{Photometric calibration to 2MASS}
\label{PhotSect}

To photometrically place our observations on the 2MASS zero-point system, we performed the following steps. First, we convolved the full grid of BT-Settl model atmospheres \citet{AGL+03} with the 2MASS\footnote{http://www.ipac.caltech.edu/2mass/releases/allsky/doc/sec3\_1b1.html} and VISTA\footnote{http://apm49.ast.cam.ac.uk/surveys-projects/vista/technical/filter-set} filter transmissions, computing the flux expected for each model in each filter. We use the [Fe/H] = 0 dex, 10\,000 K, log $g$ = 4.0 dex model as a substitute for Vega. Models with $(J - K_s)_{\rm 2MASS}$ = 0.3 to 0.4 mag were chosen for calibration for three reasons: (1) there is little variation in colour with metallicity or gravity for stars bluer than this colour; (2) there is little slope in the colour-correction term over this colour range for giant stars; and (3) there are a sufficiently-large number of observed stars in this colour range. We found the following linear relationships over this range:
\begin{eqnarray}
	(Z_{\rm VISTA} \!-\! J_{\rm 2MASS}) \!&\!=\!&\! 0.8968 \, (J \!-\! K_s)_{\rm 2MASS} - 0.0144 , \nonumber\\
	(J_{\rm VISTA} \!-\! J_{\rm 2MASS}) \!&\!=\!&\! -0.0521 \, (J \!-\! K_s)_{\rm 2MASS} + 0.0006 , \nonumber\\
	(K_{\rm VISTA} \!-\! K_{\rm 2MASS}) \!&\!=\!&\! -0.0002 \, (J \!-\! K_s)_{\rm 2MASS} - 0.0001 . \nonumber\\
\label{TransEq}
\end{eqnarray}

These relationships are valid for unextincted stars. The colour excess $E(J-K)$ is roughly half of $E(B-V)$ \citep{Rich10}, though this can for different values of $R = A_{\rm V} / E(B-V)$. The extinction towards the core of the Sgr dSph, M54, is given by \citet{Harris96} as $E(B-V) = 0.15$ mag, thus we assume $E(J-K) = 0.08$ mag. We estimate that $E(Z-J)$ is roughly a third of $E(B-V)$. While the amount of extinction does vary across the observed region by typically $\Delta E(B-V) \approx \pm 0.05$ mag \citep{SFD98}, this change is relatively small and can be neglected, imparting an error to our zero-point magnitudes of $\Delta E(J-K)$ times the gradients in Eq.\ (\ref{TransEq}).

Given $E(J-K) = 0.08 \pm 0.03$ mag, we therefore compare with stars observed by 2MASS to have 0.38 $\leq (J - K_s) \leq$ 0.48 mag. There are around 1000 such stars between $10 \leq K_s \leq 14$ mag per VISTA tile, with an average colour of $(J - K_s)_{\rm 2MASS} = 0.438 \pm 0.001$ mag in each case. Most of these stars are giants in the Galactic Bulge. For each tile, we compute the magnitude difference between the 2MASS and VISTA systems on a per-tile basis, and offset them appropriately using the transformation factors listed in Eq.\ (\ref{TransEq}). On this basis, we believe our zero points are accurate to approximately 23, 1.5 and 1 millimagnitudes for the $Z$, $J$ and $K_s$ bands, respectively, not including small errors arising from uncertainty in filter transmission curves and instantaneous variations due to weather, etc.

We finally correct our $J$- and $K_s$-band data to the 2MASS system using the following approximations:
\begin{eqnarray}
	(J_{\rm VISTA} \!\!&\!\!=\!\!&\!\! J_{\rm 2MASS}) \!-\! 0.05 \, (J \!-\! K_s)_{\rm 2MASS} + 0.0006 , \nonumber\\
	(K_{\rm VISTA} \!\!&\!\!=\!\!&\!\! K_{\rm 2MASS}). \nonumber
\label{Trans2Eq}
\end{eqnarray}
This transformation should be accurate to within 10 millimagnitudes for --0.2 $\lesssim$ $(J-K_s)$ $\lesssim$ 0.9 mag for almost all stars, with the possible exception of very cool oxygen- or carbon-rich stars.

\begin{figure}
\centerline{\includegraphics[height=0.47\textwidth,angle=-90]{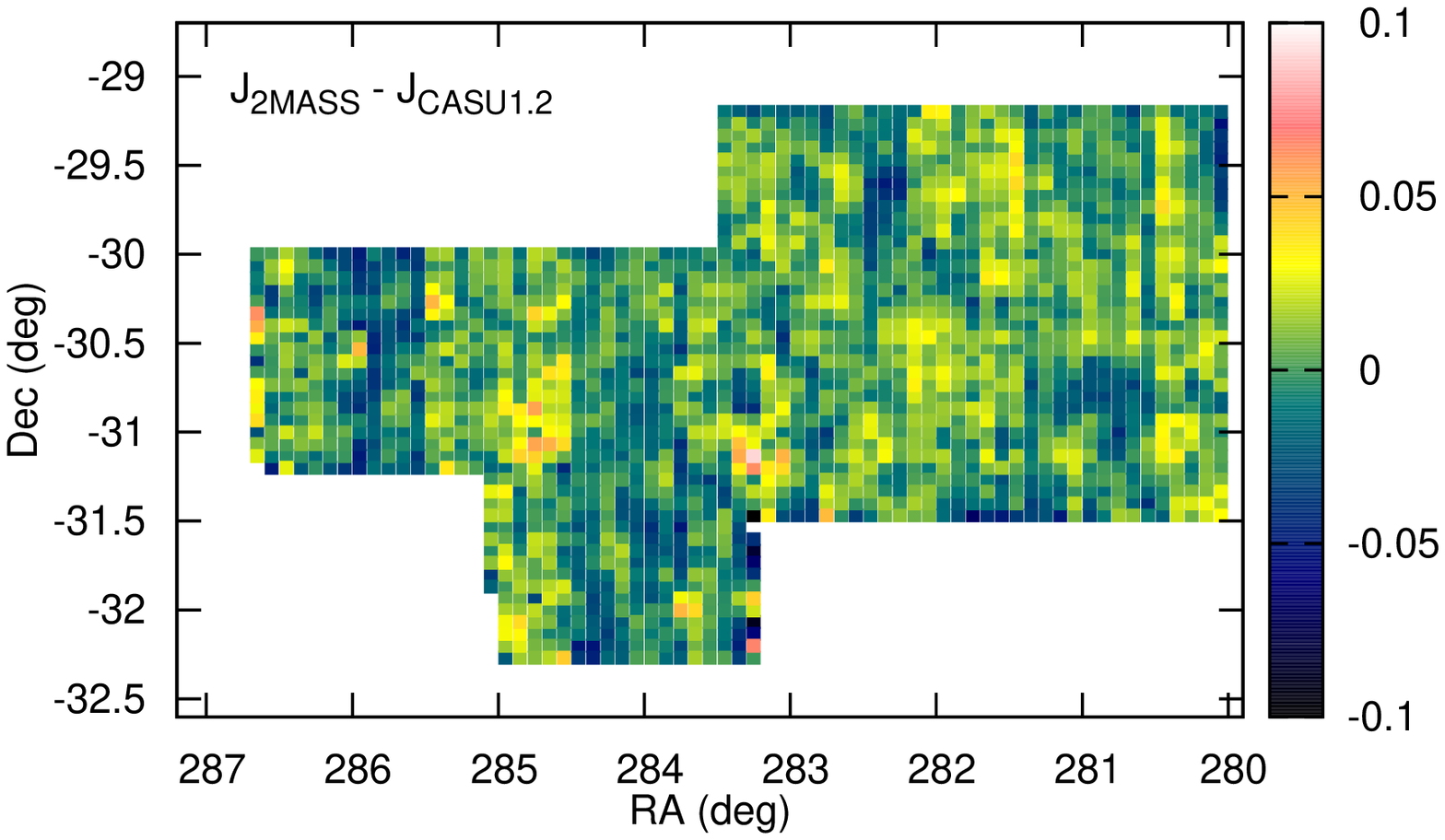}}
\vspace{-2mm}
\centerline{\includegraphics[height=0.47\textwidth,angle=-90]{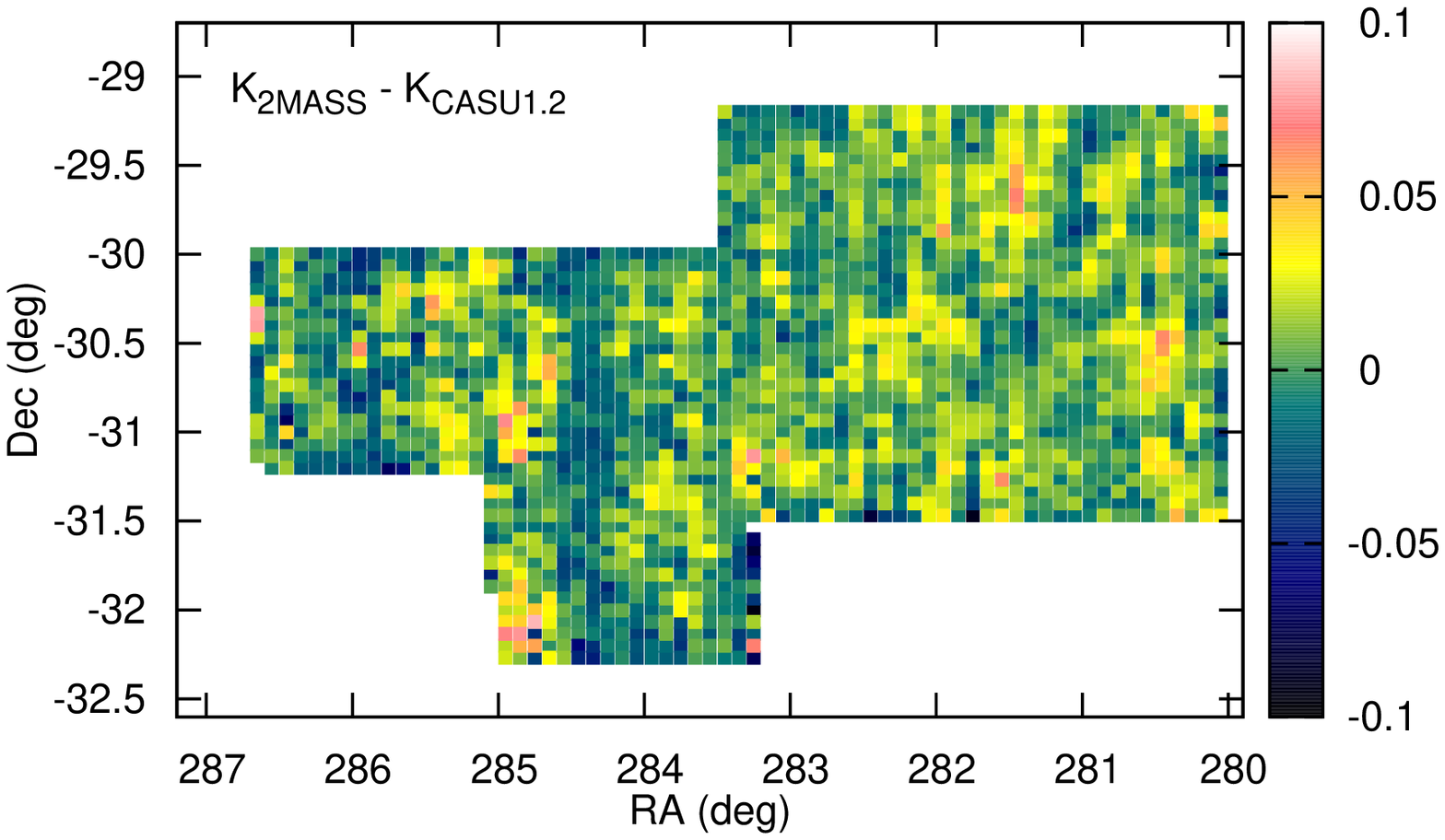}}
\caption{Mean magnitude differences per pixel between VISTA and 2MASS photometry remaining after full calibration.}
\label{DiffFig}
\end{figure}

Figure \ref{DiffFig} shows the remaining differences between the VISTA and 2MASS photometry, once all photometric calibrations have taken place. The data here are limited to those sources with an AAA data quality flag in 2MASS and a mean difference of $<$0.2 magnitudes in each band (to avoid erroneously including variable and saturated stars).

A clear variation of a few hundredths of magnitudes can be seen, causing vertical stripes in the offset maps. As both the 2MASS and VISTA surveys have a near-north--south alignment, choosing the survey which contributes most to these systematic errors is not trivial. The stripes do appear to correspond to the VISTA tile width of 1$\fdg$5 (cf.\ the 2MASS tile width is 8$\farcm$5). Examination of the VISTA data and data processing revealed no obvious errors indicating any issues in the VISTA data (M.\ Irwin, private communication).


\label{lastpage}

\end{document}